\documentclass[aps,
prd,
english,
nofootinbib,
longbibliography,
showkeys,
reprint
]{revtex4-2}

\pdfoutput=1
\usepackage{amsfonts,amsmath,amssymb}
\usepackage{graphicx}
\usepackage[utf8]{inputenc}
\usepackage{hyperref}
\usepackage{babel}
\usepackage{listings}
\usepackage{matlab-prettifier}
\usepackage{enumitem}
\usepackage{lipsum}
\newcommand\be{\begin{equation}}
\newcommand\ee{\end{equation}}
\newcommand\bea{\begin{eqnarray}}
\newcommand\eea{\end{eqnarray}}

\begin{document}

\def\rhoo{\rho_{_0}\!} %%neater subscript for rho, the disc level density function.
\def\rhooo{\rho_{_{0,0}}\!} %%neater subscript for rho, the disc level density function.
\def\G{{\widetilde\Gamma}}
\begin{flushright}
\phantom{
{\tt arXiv:2006.$\_\_\_\_$}
}
\end{flushright}

{\flushleft\vskip-1.4cm\vbox{\includegraphics[width=1.15in]{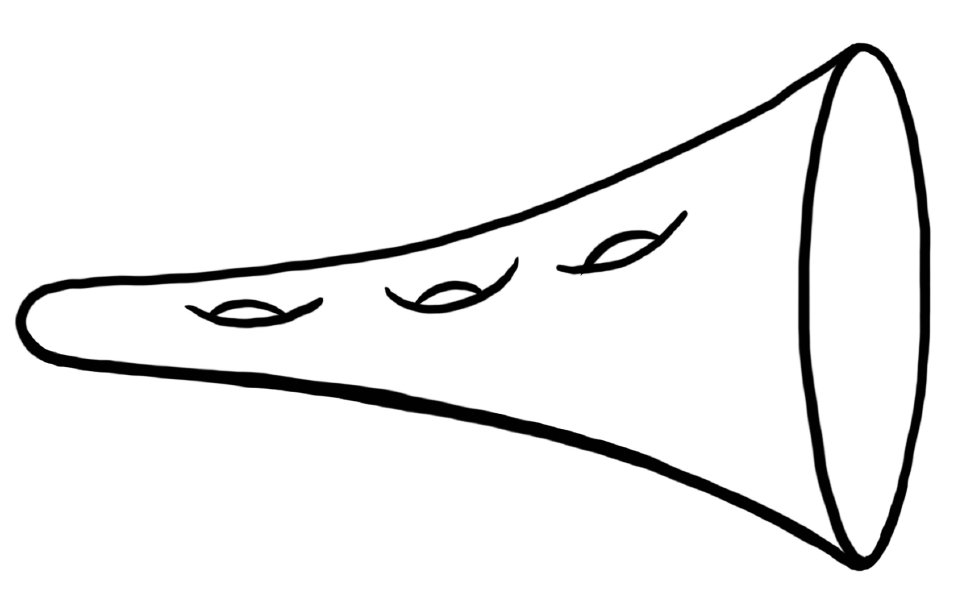}}}

\title{%{\vskip-0.4cm}
  {\Large Ramond from Random:}\\[10pt]
  Weil-Petersson Volumes for Super-Riemann surfaces\\ with  NS  Boundaries and R Punctures.}
\author{Clifford V. Johnson}
\email{cliffordjohnson@ucsb.edu}

\affiliation{Department of Physics, Broida Hall,   University of California, 
Santa Barbara, CA 93106, U.S.A.}

%\pacs{05.70.Ce,05.70.Fh,04.70.Dy}

\begin{abstract}

 The Weil-Petersson (WP) volumes of the (compactified) moduli space of ${\cal N}{=}1$ supersymmetric  Riemann surfaces with Neveu-Schwarz (NS) boundaries are frequently discussed in the literature.  Such surfaces can also have  marked points called Ramond (R) punctures,  where the superconformal structure degenerates. 
 Computing the %WP moduli space 
 volumes when these R punctures are included is more challenging for the usual differential and algebraic geometry approaches, and they are therefore  less well explored. In particular, the spectral curve describing the inclusion of  R~punctures is apparently unknown, so far. However,
the right  random matrix model approach can handle  the NS and~R sectors on an equal footing. Such a construction is presented, showing how to use a recently developed technique to readily compute many closed-form formulae for $V^{(2m)}_{g,n}(\{b_i\})$, the %Weil-Petersson 
WP
volumes for genus~$g$ with $n$ NS-boundaries of geodesic lengths $b_i$ ($i{=}1,\ldots,n$), and $2m$ R-punctures. Several striking relations between volumes (and subsectors thereof) emerge naturally in this approach. Moreover, the hitherto missing spectral curve is presented, and its use for (re-)deriving the~$V^{(2m)}_{g,n}(\{b_i\})$ is demonstrated by using  topological recursion.

\end{abstract}

%\keywords{wcwececwc ; wecwcecwc}

\maketitle

%\section{Introduction}

\section{Introduction}
%{\it Introduction}---
\label{sec:introduction}

\noindent 
Instead of deploying the heavy machinery of algebraic and differential geometry to address the matters above, this paper will follow a clear theme that has emerged in the last few decades,  beginning perhaps with the work of refs.~\cite{Witten:1989ig,Witten:1990hr,Kontsevich:1992ti}: The combined instruments of random matrix models and the integrable systems that organize their correlators give access to a remarkable amount of the geometry of interest, and at an almost preternatural level, they together unearth geometrical connections that are often well-hidden in other approaches.

In that spirit,  much of this paper will concern two related solutions  of the following ``string equation''~\cite{Morris:1990bw,Dalley:1991qg,Dalley:1992br}:
\begin{equation}
\label{eq:big-string-equation}
u{\cal R}^2-\frac{\hbar^2}2{\cal R}{\cal R}^{\prime\prime}+\frac{\hbar^2}4({\cal R}^\prime)^2=\hbar^2\Gamma^2\ ,
\end{equation}
 for the function $u(x)$. This non-linear ordinary differential equation (ODE) fully characterizes the content of a class of random matrix models to be recalled below. To unpack, a prime denotes an $x$-derivative and  each comes with an~$\hbar$ factor. Also:
 \begin{equation}
      {\cal R}{\equiv}\sum_{k=1}^\infty t_k R_k[u]{+}x\ ,
 \end{equation}
where 
$R_k[u]{=}u^k+\cdots+\# \hbar^{2k-2}u^{(2k-2)}$ is the $k$th Gel'fand-Dikii polynomial~\cite{Gelfand:1975rn} in~$u(x)$ and its $x$-derivatives, normalized here so that the purely polynomial part is unity. (Here, $u^{(m)}$ means the $m$th $x$-derivative, and intermediate terms involve products of lower derivative orders.) The first four are:
 \begin{eqnarray}
 \label{eq:GD-polynomials}
 &&R_0[u]{=}1\ ,\quad R_1[u]{=}u\ ,\quad  R_2[u]{=}u^2{-}\frac{\hbar^2}{3}u^{\prime\prime}\ , \quad \text{and}\nonumber\\
 &&R_3[u]{=}u^3{-}\frac{\hbar^2}{2}(u^\prime)^2{-} {\tiny \hbar^2}uu^{\prime\prime}{+}\frac{\hbar^4}{10}u^{\prime\prime\prime\prime}\ .
 \end{eqnarray}   
Higher $R_k[u]$ can be obtained using a recursion relation, but it is  not needed here. It will be  enough to know that at leading order in a small $\hbar$ expansion, all terms except the  $u^k$ in $R_k[u]$ can be discarded.  The  leading part of~$u(x)$ will be denoted~$u_0(x)$ in what follows.

It was recently shown in ref.~\cite{Johnson:2026twg} that the entire set of spectral correlators\footnote{The $n$ distinct $z_i$ are related to $n$ energies $E_i$ in the spectrum. The notation will be unpacked more in Section~\ref{sec:NS-boundaries}. Excellent resources explaining  properties of the $W_{g,n}$ are refs.~\cite{Eynard:2014zxa,Eynard:2016yaa}.} $W_{g,n}(\{z_i\})$ of certain matrix models in a broad class can be written simply and explicitly in terms of~$u_0(x)$ and its derivatives, with the aid of a simple operator built from~$u_0(x)$. That operator has at its origins the underlying KdV flows that $u(x,t_k)$ satisfies: ${\partial u}_{ t_k}{=}R_{k+1}^\prime[u].$
% \begin{equation}
%     \label{eq:kdv-flows}
%     \frac{\partial u}{\partial t_k}=R_{k+1}^\prime[u]\ .
% \end{equation}
Depending upon the model (and hence the particular~$u_0(x)$) the $W_{g,n}(\{z_i\})$ contain intersection theory data about a specific moduli space problem. For some cases, after an appropriate Laplace transform, these data are  Weil-Petersson volumes $V_{g,n}(\{b_i\})$~\cite{Wolpert1983SymplecticGeometry,Mirzakhani2007WPVolumesIntersection,Mirzakhani:2006fta,Eynard:2007fi,Wolpert2010WeilPetersson}.
Ref.~\cite{Johnson:2026twg}'s formulae will play a role below, as will some remarkable closed form results for $W_{g,n}$ that arise in the supersymmetric case.

String equation~(\ref{eq:big-string-equation}) was first discovered and studied extensively initially in refs.~\cite{Morris:1990bw,Dalley:1991qg,Dalley:1992br},  arising in  ``double scaling limits'' (DSL)~\cite{Gross:1990vs,Gross:1990aw,Douglas:1990ve,Brezin:1990rb} of  multicritical Wishart-type random matrix models at large $N$. It  was also shown~\cite{Dalley:1992br} to map to the physics of double-scaled unitary matrix model in external fields~\cite{Periwal:1990gf,Periwal:1990qb,Gross:1980he,Wadia:1981rb,Gross:1991aj,Brezin:1980rk}. (This will be returned to below). The matrix model is in the ($\boldsymbol{\alpha}$,$\boldsymbol{\beta}){=}(2\Gamma+1,2)$) in the Altland-Zirnbauer~\cite{Altland:1997zz} classification. 
 The parameter $\Gamma$ plays various roles depending upon context. In the Wishart setup it counts how rectangular the matrices are (see also refs.~\cite{Myers:1991akt,Lafrance:1993wy}). Alternatively, it counts  threshold bound states (solitons) in an associated integrable system~\cite{Carlisle:2005mk,Carlisle:2005wa}, while in a stringy setting it can count background D-branes~\cite{Dalley:1992br,Johnson:1994vk}, R-R flux~\cite{Klebanov:2003wg}. In more recent settings, it has been shown to naturally count BPS states in JT gravity with  extended supersymmetry~\cite{Johnson:2023ofr,Johnson:2024tgg,Johnson:2025oty}. In the present paper, as will  be discussed presently,  $\Gamma$ will count insertions of a certain kind of boundary on super-Riemann surfaces called a ``Ramond puncture''. 
 
 The  striking ease with which $\Gamma$  can be incorporated {\it via} the string equation has already been remarked upon as potentially useful for describing such insertions~\cite{Johnson:2019eik,Johnson:2020heh,Johnson:2026twg}, and this paper will confirm and elaborate on  those observations by then  {\it explicitly} constructing the  random matrix model in which Ramond (R) punctures are on an equal footing with their Neveu-Schwarz (NS) boundary counterparts ({\bf Section~\ref{sec:Ramond-from-Random}}). This entire approach has the happy consequence of being useful for swiftly generating several closed-form formulae for the Weil-Petersson volumes  $V^{(2m)}_{g,n}(\{b_i\})$ of the compactified moduli space of  genus~$g$ super-Riemann surfaces with~$n$ NS~boundaries (with lengths $\{b_i\}$) and $2m$ R~punctures, using the new methods of ref.~\cite{Johnson:2026twg}. They confirm some low genus formulae derived in the mathematical literature using other methods by Norbury\cite{Norbury:2020vyi,norbury2024superweilpeterssonmeasuresmoduli}, readily provide many more, and unearth many interesting relations among them ({\bf Section~\ref{sec:Gamma-fun}}). Missing from the literature to date has been a spectral curve from which such volumes can be computed using the techniques of topological recursion~\cite{Chekhov:2006vd,Eynard:2007kz}. This paper will provide that spectral curve ({\bf Section~\ref{sec:spectral-curve}}), and demonstrate in various ways that it produces the volumes of interest.

For the more mathematically cautious, it is perhaps worth noting that the explorations of the solutions of the string equation~(\ref{eq:big-string-equation}) (and hence the results that follow) and the many connections we shall henceforth make to the $V^{(2m)}_{g,n}$, are firmly underpinned by an important long-known connection that has been re-emphasized (and deeply mathematically explored) in recent times. As already mentioned, $u(x;t_k)$ obeys the KdV hierarchy of flows. The string equation can be thought of as a particular set of initial conditions for those flows that are in general different from those pertaining to bosonic models.\footnote{In the language of Virasoro constraints~\cite{Dijkgraaf:1991rs} the focus is on the  scale invariant sector's solutions~\cite{Dalley:1991vr,Dalley:1992br} rather than the translationally invariant ones: The string equation is $L_0$ instead of~$L_{-1}$.} On the other hand, it is often stated that the relevant integrable structure for the ${\cal N}{=}1$ supersymmetric context is the Brezin-Gross-Witten (BGW)
 hierarchy of flows, which first emerged in the context of random unitary matrix models~\cite{Brezin:1980rk,Gross:1980he,Wadia:1981rb,Periwal:1990gf,Periwal:1990qb}. In fact, it was explicitly proven very long ago in ref.~\cite{Dalley:1992br} that the solutions of  string equation~(\ref{eq:big-string-equation}) are in one-to-one correspondence with those of the ``Painlev\'e~II'' hierarchy of equations that arise in the BGW-underpinned models. Put differently, the string equation defines a $\tau$-function of the BGW hierarchy, and as we shall review, turning on the parameter (called $\Gamma$) that controls the Ramond sector is a simple and smooth (well-understood) deformation away from the purely NS case (reviewed next in Section~\ref{sec:NS-boundaries}). 
 
 Recent  work by   Alexandrov and Norbury~\cite{norbury2024superweilpeterssonmeasuresmoduli,Alexandrov:2024kuj}, building on and mathematically expanding Stanford and Witten's work in ref.~\cite{Stanford:2019vob}, has 
centered this deformation (and the relevance to BGW) in this R-puncture context, providing many powerful proofs using differential and algebraic geometry that can taken as additional assurance that what is to follow is on a firm mathematical foundation.
As we shall soon see, the string equation approach will yield many new results that are of  mathematical interest. It will be interesting to see how some of them can be understood in the geometrical framework discussed in those papers.

  As should be clear from all that has been covered in this section, the many results  will simply flow from identifying the relevant solutions for $u(x)$ of  equation~(\ref{eq:big-string-equation}), so we will turn to discussing solutions next, as well as what to do with them to construct quantities of interest. 
  
 The parameter~$\hbar$
 is the  $1/N$ expansion parameter (renormalized after the DSL) of the random matrix model and in the geometric interpretation that emerges from these models, the expansion organizes quantities by topology, following from the usual 't Hooftian construction~\cite{'tHooft:1973jz,Brezin:1978sv,Bessis:1980ss}. The key function $u(x)$ has the expansion:
 \begin{equation}
u(x)=u_0(x)+\sum_{g=1}^\infty u_{2g}\hbar^{2g}+\cdots
\label{eq:u-expansion}
 \end{equation}
 where the ellipsis denotes non-perturbative parts.
 
 Solutions of the equation can be specified by describing the leading behaviour for $u_0(x)$ on all of $x$.  As a boundary value problem, this provides the needed boundary conditions in the far $x{<}0$ and $x{>}0$ directions. A simple iterative procedure can develop expressions for the $u_{2g}(x)$ in terms of the initial $u_0(x)$ in a given region. Moreover, the equation is known to  supply rich non-perturbative information that has been explored extensively in this context in {\it e.g.} refs.~\cite{Johnson:2019eik,Johnson:2020exp,Johnson:2021owr,Johnson:2022wsr}.

 \section{${\cal N}{=}1$ supersymmetry and Neveu-Schwarz boundaries}
 \label{sec:NS-boundaries}
\begin{figure}[t]
    \centering
    \includegraphics[width=0.99\linewidth]{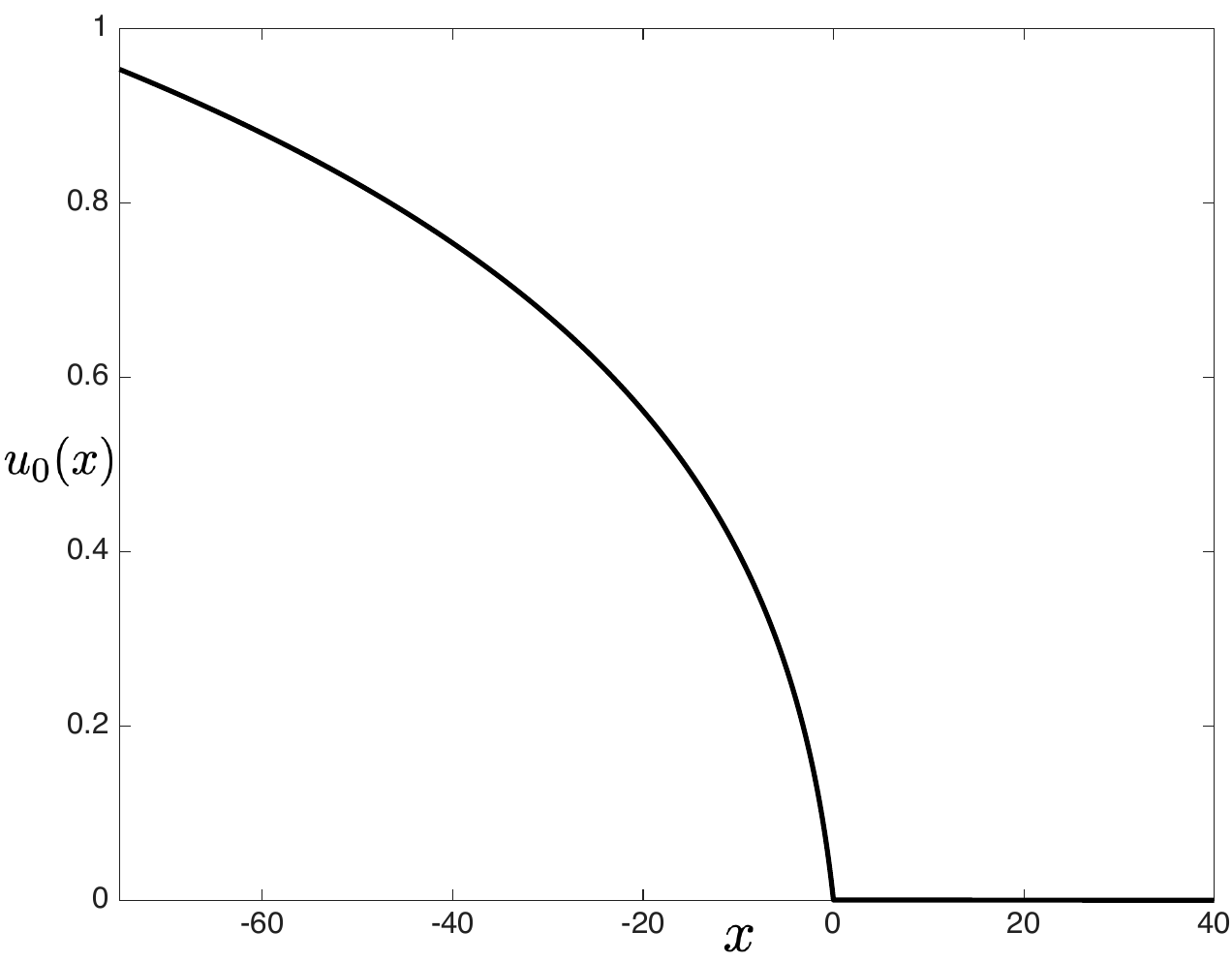}
    \caption{A plot of $u_0(x)$ for the purely NS  case, consisting of $u_0(x)$ solving equation~(\ref{eq:leading_N=1-string-equation}), for $x\leq0$ and $u_0(x)=0$ for $x>0$.}
    \label{fig:u-functions-no-Gamma}
\end{figure}

 The first solution we shall discuss has leading part~$u_0(x)$ defined by setting $\hbar{=}0$ in~(\ref{eq:big-string-equation}) and solving:
 \begin{equation}
     u_0{\cal R}_0^2=0\ ,\quad \text{with}\quad {\cal R}_0\equiv\sum_{k=1}^\infty t_k u_0^k+x\ .
     \label{eq:classical-one}
 \end{equation}
First constructed in ref.~\cite{Johnson:2020heh}, it has a piecewise description. For $x>0$ the solution is $u_0=0$ while for $x<0$ the solution is ${\cal R}_0{=}0$. With the following  values for~$t_k$:
\begin{equation}
    t_k=\frac{\pi^{2k}}{(k!)^2}\ ,
\end{equation} the positive part of the function $u_0$ therefore solves:
\begin{equation}
I_0(2\pi\sqrt{u_0})-1+x=0\ ,
\label{eq:leading_N=1-string-equation}
\end{equation}
where $I_0$ is the modified Bessel function. The solution is plotted  in figure~\ref{fig:u-functions-no-Gamma}.

The leading spectral density that follows from this is constructed from the following integral representation:
\begin{equation}
    \label{eq:integral-representation}
    \rho_0(E)=\frac{1}{2\pi\hbar}\int_{-\infty}^\mu\frac{\Theta(E-u_0(x)) dx}{\sqrt{E-u_0(x)}}\ ,
\end{equation}
where here $\mu\equiv t_0=1$, giving:

\begin{equation}
\rho_0(E)=\frac{\cosh(2\pi\sqrt{E})}{2\pi\hbar\sqrt{E}}\ .
\label{eq:N-1-density}
\end{equation}
This is the super-Schwarzian spectral density~\cite{Mertens:2017mtv,Stanford:2017thb}, which played a central role in Stanford and Witten's studies of ${\cal N}{=}1$ JT supergravity~\cite{Stanford:2019vob}.
The full solution of the equation for the cases of the models defines by $\Gamma{=}0$, and $\Gamma{=}\pm\frac12$ were explored in ref.~\cite{Johnson:2020exp}.

Perturbation theory in the random matrix model is built from  $u(x)$ evaluated at $x=\mu=1$.  The appropriate expansion in this positive $x$ regime  has the following leading terms for the case of $\Gamma{=}0$:
\begin{eqnarray}
\label{eq:perturbative-u}
u(x)
&=&0
-\hbar^2\frac{1}{4x^2}
+\hbar^4\frac{9t_1}{8x^5}
\\
&&\hskip1.82cm
+\hbar^6\frac{9}{64}
\left[
\frac{50t_2}{x^7}
-\frac{147t_1^2}{x^8}
\right]
+\cdots \ .\nonumber
\end{eqnarray}
These data can all be translated into corrections to $\rho(E)$ or correlation  functions of multiple copies of it, denoted:
\begin{equation}
    {\widetilde W}_{g,n}(\{E_i\})=\langle \rho(E_1)\cdots\rho(E_n)\rangle_g\ ,\end{equation}
    which comes at order $\hbar^{\chi}=\hbar^{2g-2+n}$. How to do this efficiently in the matrix model was recently explained in ref.~\cite{Johnson:2026twg}. Our interest will be in the quantities obtained by sending $E_i\to-E_i$ and  defining the  uniformizing coordinates ${\hat z}_i$ {\it via}  ${\hat z}_i^2=-E_i$. (It is important that these ${\hat z}_i$ have been defined  with a hat (as compared to standard practice). This will  avoid a notational confusion with a distinct set of local coordinates $z_i$ to appear shortly.)  
    
    After multiplying by a Jacobian $\prod_{i=1}^n(-2{\hat z}_i)$
the result is the set of correlators $W_{g,n}(\{{\hat z}_i\})$. They are in fact  Laplace transforms:
\begin{equation}
    W_{g,n}(\{\hat z_i\}) = \int \prod b_i db_i V_{g,n}(\{b_i\}){\rm e}^{-b_i{\hat z}_i}\ ,
\end{equation}
of the $V_{g,n}(\{b_i\})$ which are the super-Weil-Petersson volumes first studied in ref.~\cite{Stanford:2019vob}. There, a generalization of Mirzakhani's recursion for ordinary WP volumes~\cite{Mirzakhani:2006fta} was derived for these.
Note that defining the following algebraic curve  $(X(\hat z),Y(\hat z))$  by: 
\begin{eqnarray}
    -E &=& X(\hat z) = \hat z^2\ ,\\
    \pi\hbar i\rho_0(E) &=& Y(\hat z) = -\frac{\cos(2\pi \hat z)}{2\hat z}\ .
    \label{eq:spectral-curve-NS}
\end{eqnarray}
yields a spectral curve that can be used to define the $W_{g,n}(\{\hat z_i\})$ {\it via} the topological recursion framework, in an analogous way to what was done for ordinary WP volumes in ref.~\cite{Eynard:2007fi},
where instead (in our conventions) $Y(z){=}\sin(2\pi z)/4\pi$.\footnote{The minus sign in $Y({\hat z})$ in (\ref{eq:spectral-curve-NS}) is a bit unusual compared to some choices in the literature. It results in a $(-)^n$ in front of the ${\cal N}{=}1$ supersymmetric $W_{g,n}$ and $V_{g,n}$ relative to other choices. However, it will fit nicely with the various conventions of this paper and ref.~\cite{Johnson:2026twg}, and also the conventions of ref.~\cite{Stanford:2019vob}.}

\section{The role of parameter $\Gamma$}
\label{sec:Gamma-fun}
Now comes a key point. It is  easy to generate $W_{g,n}$'s with non-zero $\Gamma$ using the methods of ref.~\cite{Johnson:2026twg}. The more general expansion of $u(x)$ begins:\footnote{In fact, for $\Gamma{=}\pm\frac12$ the expansion vanishes to all orders in perturbation theory. In fact, for these special values the models are non-orientable~\cite{Stanford:2019vob}.\label{fn:special-models}}
\begin{widetext}
\begin{eqnarray}
\label{eq:perturbative-u-Gamma}
&&u(x)=0+\hbar^2\left(\frac{\Gamma^2-\frac14}{x^2}\right)+\hbar^4\left(2t_1\frac{\left(\Gamma^2-\frac14\right)\left(\Gamma^2-\frac94\right)}{x^5}\right)\nonumber\\&&\hskip3.9cm +\hbar^6\left(\Gamma^2-\frac14\right)\left(\Gamma^2-\frac94\right)
%\times\\
%&&\hskip2.3cm
\left[
\frac{7t_1^2}{x^8}\left(\Gamma^2-\frac{21}{4}\right)
-\frac{2t_2}{x^7}\left(\Gamma^2-\frac{25}{4}\right)
\right]+\cdots\ .
\end{eqnarray}
\end{widetext}

Since $\Gamma$ is subleading in the expansion of the $u(x)$ it does not appear in the leading spectral density, and hence does not appear in the spectral curve. So as it stands,~$\Gamma$ is invisible as far as (standard) topological recursion is concerned, and so the natural volumes (and $W_{g,n}$) accessible with the spectral curve~(\ref{eq:spectral-curve-NS}) are those with NS boundaries only. One of the outcomes of this paper will be to show how to deform this spectral curve to get a new one that incorporates $\Gamma$.

On the other hand, the recent method of ref.~\cite{Johnson:2026twg},  takes as direct input the string equation solution for $u(x)$ to build $W_{g,n}$, including the details of the higher order corrections beyond $u_0$. Another output of that method  was a set of  swift proofs of some striking closed-form expressions for the $W_{g,n}$ for  $g{=}1,2,3$ (first derived by Norbury~\cite{Norbury:2020vyi}), and also their generalization to include $\Gamma$.  Moreover the approach  also provided a method  for how to simply derive new closed-form expressions for any~$g$ (with the new example of $g=4$ done explicitly.) 
\begin{widetext}
 Here are the  expressions from ref.~\cite{Johnson:2026twg}, which will be useful shortly:
\begin{eqnarray}
    &&W_{1,n}(\{{\hat z}_i\};\Gamma) = (-1)^{n+1}\frac{(n-1)!}{2}\left(\Gamma^2-\frac14\right)\prod_{i=1}^n
 \frac{1}{{\hat z}_i^2}\ ,\label{eq:W1n-special}\\
    &&W_{2,n}(\{{\hat z}_i\};\Gamma) =(-)^n\frac{(n+1)!}{6\cdot4^2}\left(\Gamma^2-\frac14\right)\left(\Gamma^2-\frac94\right)\prod_{i=1}^n \frac{1}{{\hat z}_i^2}
  \left\{
4t_1(n+2)+6\sum_{i=1}^n\frac{1}{{\hat z}_i^2} \right\}
\ ,
\label{eq:W2n-special}  \\
    &&W_{3,n}(\{{\hat z}_i\};\Gamma) =(-1)^{n+1}\frac{ (n+3)!}{5760}\,
\Bigl(\Gamma^2-\frac94\Bigr)\Bigl(\Gamma^2-\frac14\Bigr)\prod_{i=1}^{n}\frac{1}{{\hat z}_i^2}
\Bigg[
4(n+4)
\Bigl(
(n+5)\Bigl(\Gamma^2-\frac{21}{4}\Bigr)t_1^2
-2\Bigl(\Gamma^2-\frac{25}{4}\Bigr)t_2
\Bigr)
\nonumber\\
&&\qquad\qquad
+12(n+4)\Bigl(\Gamma^2-\frac{21}{4}\Bigr)t_1\,\sum_{i=1}^{n}\frac{1}{{\hat z}_i^2}
+15\Bigl(\Gamma^2-\frac{25}{4}\Bigr)\sum_{i=1}^{n}\frac{1}{{\hat z}_i^4}
+18\Bigl(\Gamma^2-\frac{21}{4}\Bigr)\sum_{1\le i<j\le n}\frac{1}{{\hat z}_i^2 {\hat z}_j^2}
\Bigg]\ ,
    \label{eq:W3n-special}
    \\
    &&W_{4,n}(\{{\hat z}_i\};\Gamma) = \text{(see equation~(\ref{eq:W4n-general}))}\label{eq:W4n-special}
\end{eqnarray}
\end{widetext}
where $W_{4,n}(\{{\hat z}_i\};\Gamma)$ has been placed in Appendix~\ref{app:W4n-closed-formula} due to its length, and $W_{0,n}(\{{\hat z}_i\})=0$ for $n>2$ following from the exact vanishing of $u_0(x)$ for $x>0$.

 Considering it as a parameter inserted into the $\Gamma{=}0$ theory, what  is $\Gamma$ doing for us? Long ago~\cite{Dalley:1992br} it was noticed that it resembles a closed string insertion in the $x>0$ regime of perturbation theory of the string equation, while in the $x<0$ regime it has a natural interpretation as an open string insertion: $\hbar\Gamma$ introduces boundaries (D-branes) with Chan-Paton factor $\Gamma$. In the $x>0$ regime, relevant for us here, it appears only as $(\hbar\Gamma)^2$ and powers thereof and so is naturally a closed-string type pointlike insertion that evidently can also be considered a collapsed boundary or puncture. (In more modern parlance,  the string equation describes a geometric transition in which the D-branes dissolve and are replaced by closed string (flux) insertions.)  A direct connection to the R-R sector was made when  the string equation was connected~\cite{Klebanov:2003wg} to the type~0A minimal superstring theory. Moreover ref.~\cite{Stanford:2019vob}  recognized that a $(2\Gamma+1,2)$-type random matrix model should allow for the description of Ramond punctures. Refs.~\cite{Johnson:2019eik,Johnson:2020heh} made this explicit at the level of string equation solutions for $\Gamma=0$, and recently ref.~\cite{Johnson:2026twg} took it further. 

 The point is that $\Gamma$ needs to be large, scaling with~$N$ in the original random matrix model. So as $\hbar\to0$ the combination ${\widetilde\Gamma}=\hbar\Gamma$ should be held fixed. It is this combination that will count Ramond insertions that can be made visible in the leading description of the  geometry. %, and ultimately the spectral curve deformation will be controlled by this parameter, as we will see.

So although the NS and R sectors are not on the same footing in the model described so far, ref.~\cite{Johnson:2026twg} showed (see the Discussion section there) that one can re-organize the results to recover some already known formulae~\cite{Norbury:2020vyi} for the Weil-Petersson volumes (or Laplace transform) with R-punctures present. By rewriting   the formulae~(\ref{eq:W1n-special})--(\ref{eq:W3n-special}) in terms of $\G$, each replacement of  $\Gamma^2$ by ${\widetilde\Gamma}^2/\hbar^2$ in a $W_{g,n}$ loses two powers of $\hbar$, and so ${\widetilde\Gamma}^{2m}$ now multiplies what might be better called $W^{(2m)}_{g-m,n}$, where the superscript counts how many Ramond boundary insertions there are on the Riemann surfaces. 

\begin{widetext}
We can continue the job here, yielding (for example) non-zero genus zero correlators:
\begin{eqnarray}
 &&W_{0,n}^{(2)}=(-1)^{n+1}\frac{(n-1)!}{2}\prod_{i=1}^n
 \frac{1}{{\hat z}_i^2}\ ,
 \label{eq:W0n2-special}
\\
  &&W_{0,n}^{(4)}(\{{\hat z}_i\})
=(-)^n\frac{(n+1)!}{6\cdot4^2}\prod_{i=1}^n \frac{1}{{\hat z}_i^2}
  \left\{
4t_1(n+2)+6\sum_{i=1}^n\frac{1}{{\hat z}_i^2} \right\}
\ ,
\label{eq:W0n4-special}
\\
&&W^{(6)}_{0,n}(\{\hat z_i\})
={}
(-1)^{n+1}\frac{(n+3)!}{5760}
\prod_{i=1}^{n}\frac{1}{\hat z_i^2}
\Bigg[
4(n+4)(n+5)t_1^2
-8(n+4)t_2
+12(n+4)t_1
\sum_{i=1}^{n}\frac{1}{\hat z_i^2}
\nonumber\\
&&\hspace{11.0cm}
+15
\sum_{i=1}^{n}\frac{1}{\hat z_i^4}
+18
\sum_{1\le i<j\le n}
\frac{1}{\hat z_i^2\hat z_j^2}
\Bigg] \ ,
\label{eq:W0n6-special}
\end{eqnarray} 
as well as:
\begin{align}
W^{(8)}_{0,n}
={}&
(-1)^n (n+5)!\,
\prod_{i=1}^n\frac{1}{\hat z_i^2}
\Bigg\{
(n+6)
\left[
\frac{(n+7)(n+8)}{241920}\,t_1^3
-\frac{n+7}{40320}\,t_1t_2
+\frac{1}{40320}\,t_3
\right]
\nonumber\\
&\hspace{2.0cm}
+(n+6)
\left[
\frac{n+7}{53760}\,t_1^2
-\frac{1}{26880}\,t_2
\right]
\sum_{i=1}^n\frac{1}{\hat z_i^2}
%\nonumber\\
%&\hspace{6.3cm}
+\frac{(n+6)t_1}{21504}
\sum_{i=1}^n\frac{1}{\hat z_i^4}
+\frac{(n+6)t_1}{17920}
\sum_{1\leq i<j\leq n}
\frac{1}{\hat z_i^2\hat z_j^2}
\nonumber\\
&\hspace{5.0cm}
+\frac{1}{18432}
\sum_{i=1}^n\frac{1}{\hat z_i^6}
+\frac{1}{14336}
\sum_{\substack{1\leq i,j\leq n\\ i\neq j}}
\frac{1}{\hat z_i^4\hat z_j^2}
+\frac{3}{35840}
\sum_{1\leq i<j<k\leq n}
\frac{1}{\hat z_i^2\hat z_j^2\hat z_k^2}
\Bigg\}\ .
\label{eq:W0n8-special}
\end{align}

\noindent
Some genus one results are:
 \begin{eqnarray}
 &&W_{1,n}^{(0)}(\{{\hat z}_i\})=(-1)^{n+1}\frac{(n-1)!}{2}\left(-\frac14\right)\prod_{i=1}^n
 \frac{1}{{\hat z}_i^2}\ ,
 \label{eq:W1n0-special}\\
 &&W_{1,n}^{(2)}(\{{\hat z}_i\})
=(-)^n\frac{(n+1)!}{6\cdot4^2}\left(-\frac{10}{4}\right)\prod_{i=1}^n \frac{1}{{\hat z}_i^2}
  \left\{
4t_1(n+2)+6\sum_{i=1}^n\frac{1}{{\hat z}_i^2} \right\}
\ ,
\label{eq:W1n2-special}\\
&&W_{1,n}^{(4)}(\{{\hat z}_i\})
={}
(-1)^{n+1}\frac{(n+3)!}{5760}
\prod_{i=1}^{n}\frac{1}{\hat z_i^2}
\Bigg[
-31(n+4)(n+5)t_1^2
+70(n+4)t_2
\nonumber\\
&&\hspace{6.0cm}
-93(n+4)t_1
\sum_{i=1}^{n}\frac{1}{\hat z_i^2}
-\frac{525}{4}
\sum_{i=1}^{n}\frac{1}{\hat z_i^4}
-\frac{279}{2}
\sum_{1\le i<j\le n}
\frac{1}{\hat z_i^2\hat z_j^2}
\Bigg] .
\label{eq:W1n4-special}
\end{eqnarray}
Finally we extract some  genus two results:
\begin{align}
W^{(0)}_{2,n}
={}&
(-1)^n (n+1)!\,
\prod_{i=1}^n\frac{1}{\hat z_i^2}
\left[
\frac{3(n+2)}{128}\,t_1
+\frac{9}{256}
\sum_{i=1}^n\frac{1}{\hat z_i^2}
\right]\ ,
\\[6pt]
W^{(2)}_{2,n}
={}&
(-1)^{n+1} (n+3)!\,
\prod_{i=1}^n\frac{1}{\hat z_i^2}
\Bigg[
\frac{73}{7680}(n+4)(n+5)t_1^2
-\frac{259}{11520}(n+4)t_2
+\frac{73}{2560}(n+4)t_1
\sum_{i=1}^n\frac{1}{\hat z_i^2}
\nonumber\\
&\hspace{9.0cm}
+\frac{259}{6144}
\sum_{i=1}^n\frac{1}{\hat z_i^4}
+\frac{219}{5120}
\sum_{1\leq i<j\leq n}
\frac{1}{\hat z_i^2\hat z_j^2}
\Bigg]\ .
\end{align}
The remainder ($W^{(6)}_{1,n}$, $W^{(4)}_{2,n}$, $W^{(0)}_{3,n}$, $W^{(2)}_{3,n}$, and $W^{(0)}_{4,n}$) can be extracted as well, but the expressions get successively longer, so they can be readily  be deduced from  equations~(\ref{eq:W1n-special})--(\ref{eq:W4n-special}) as needed.
As one of many nice checks one can do on this we can gather all the disc $(g{=}0,n{=}1)$ results:
\begin{align}
W^{(2)}_{0,1}&=\frac{1}{2{\hat z}^2}\ ,
\quad W^{(4)}_{0,1}=-\frac{t_1}{4{\hat z}^2}-\frac{1}{8{\hat z}^4}\ ,
\quad W^{(6)}_{0,1}=\left(\frac12t_1^2-\frac16t_2\right)\frac{1}{{\hat z}^2}
+\frac{t_1}{4{\hat z}^4}
+\frac{1}{16{\hat z}^6}\ ,
\\
\text{and}\quad W^{(8)}_{0,1}&=\left(-\frac32t_1^3+t_1t_2-\frac18t_3\right)\frac{1}{{\hat z}^2}
+\left(-\frac34t_1^2+\frac{3}{16}t_2\right)\frac{1}{{\hat z}^4}
-\frac{15}{64}\frac{t_1}{{\hat z}^6}
-\frac{5}{128{\hat z}^8} \ .
\label{eq:disc-W08}
\end{align}
After setting $t_k{=}\pi^{2k}/(k!)^2$ this gives, for $W_{0,1}^{\rm R}=\sum_m\G^{2m}W^{(2m)}_{0,1}$:
\begin{align}
W_{0,1}^{\rm R}
={}&
\Bigg[
\frac{\widetilde\Gamma^2}{2{\hat z}^2}
-\widetilde\Gamma^4
\left(
\frac{\pi^2}{4{\hat z}^2}
+\frac{1}{8{\hat z}^4}
\right)
+\widetilde\Gamma^6
\left(
\frac{11\pi^4}{24{\hat z}^2}
+\frac{\pi^2}{4{\hat z}^4}
+\frac{1}{16{\hat z}^6}
\right)
-\widetilde\Gamma^8
\left(
\frac{361\pi^6}{288{\hat z}^2}
+\frac{45\pi^4}{64{\hat z}^4}
+\frac{15\pi^2}{64{\hat z}^6}
+\frac{5}{128{\hat z}^8}
\right)
\Bigg]
+O(\widetilde\Gamma^{10})\ ,
\label{eq:Ramond-disc-expansion}
\end{align}
which is precisely the output of the disc relation that Norbury proved in ref.~\cite{norbury2024superweilpeterssonmeasuresmoduli}:
\begin{equation}
\label{eq:Norbury-disc-relation}
F(\hat z)
=
\frac{s^2}{2\hat z^2}
+
\left[
\frac{F(\hat z)^2}{2\cos(2\pi \hat z)}
\right]_{\hat z=0}\ ,
\end{equation}
if we identify $W_{0,1}^{\rm R}(\hat z,\G)=-F(\hat z,s)|_{s^2=-\widetilde\Gamma^2}$ ({\it i.e.,}  $s^2$ there plays the role of our $-\widetilde\Gamma^2$), and $[\cdots]_{\hat z=0}$ means taking the principal part (inverse powers) of the
Laurent expansion at $\hat z{=}0$. We will soon see that there is a quite different closed form way, {\it via} the string equation approach, of yielding the content of~(\ref{eq:Ramond-disc-expansion})---see equations~(\ref{eq:exact-density-i0}) and~(\ref{eq:endpoint-equation})---and that form will immediately yield the  disc spectral curve in Section~\ref{sec:spectral-curve-soft}, and particularly Section~\ref{sec:spectral-curve-hard}. 
\end{widetext}
    
There is another striking thing being revealed  by these general formulae about the $W^{(2m)}_{g,n}$  (and hence, by Laplace transform, the $V^{(2m)}_{g,n}$): There are many structural similarities shared among them. For example, $W^{(2)}_{0,n}=-\frac14 W^{(0)}_{1,n}$, and a host of similar relations that skip one genus following from the fact that there is an overall $\Gamma^2-\frac14$ in front of every $W_{g,n}(\Gamma)$ in equations~(\ref{eq:W1n-special})--(\ref{eq:W4n-special}), which in turn has its origins in the expansion for $u(x)$ in equation~(\ref{eq:perturbative-u-Gamma}).\footnote{Such a relation can perhaps be interpreted as the reason  why the $\Gamma=\pm\frac12$ models have perturbation theory vanishing to all orders: NS and R contributions appear with just the right relative weights to cancel. See footnote~\ref{fn:special-models}.} There are analogues of these relations resulting from   vanishings for other half integer $\Gamma$ too. These all would be interesting to understand  in a geometrical approach. 
 Really intriguing however are the occurrence of special vanishings within substructures in the volume formulae, and even at non-half integer $\Gamma$, such as at $\Gamma^2=\frac{21}{4}$ as can be seen in equation~(\ref{eq:W3n-special}), or  
$\Gamma^2{=}\frac{83}{12}$, $\Gamma^2{=}\frac{29}{4}$, and so on, as seen in (\ref{eq:W4n-general}). Quite generally, a vanishing at a special value of $\Gamma$ says that there are relations between certain weighted contributions in the sum. So
for  example, in the genus~$3$ formula~(\ref{eq:W3n-special}), the substructure built from terms involving 
$t_1^2$, $t_1\sum_i \hat z_i^{-2}$, and
$\sum_{i<j}\hat z_i^{-2}\hat z_j^{-2}$ has the factor
$\Gamma^2{-}\frac{21}{4}$. The special vanishing at $\Gamma^2{=}\frac{21}{4}$
implies  a vanishing linear combination of the parts of $W^{(0)}_{3,n},W^{(2)}_{2,n},W^{(4)}_{1,n},$ and $W^{(6)}_{0,n}$ that have been projected to that subsector. This is a clear set of predictions (and many others are easily generated) about the relation between volumes (and substructures thereof) that deserves deeper understanding.

Moving on from this, the goal now is to find the definition of the  matrix model that treats the NS and R sectors on the same footing, rather than to have to take the above limit. In other words: What is the $u_0(x)$ for such a model,  the corresponding spectral density $\rho_0(E)$, and hence the equivalent spectral curve $(X(\hat z),Y(\hat z))$? 
That $u_0(x)$ corresponds to a re-organization of the entire perturbative series~(\ref{eq:perturbative-u-Gamma}), so that the leading part is
%\begin{equation}   
$u_0(x){=}{{\widetilde\Gamma}^2}/{x^2}+\cdots 
$
%\end{equation}
as $x\to\infty$.
Leading solutions of the string equation that incorporate non-zero $\Gamma$ in this way were first considered in ref.~\cite{Johnson:2023ofr} in the context of ${\cal N}{=}2$ JT supergravity~\cite{Turiaci:2023jfa} where~$\Gamma$ counts BPS states. 

There is no reason why solutions of that sort can't apply for the ${\cal N}{=}1$ case here. The result would be a non-zero leading classical piece  $u_0(x)$ from which, using the general formulae of ref.~\cite{Johnson:2026twg}, all correlators $W_{g,n}^{(2m)}$ can be derived directly, by simply reading off the coefficients of powers of~${\widetilde\Gamma}$. This would be a description of the full matrix model, rather than taking a limit on the results given above. That there are non-vanishing genus zero correlators $W^{(2m)}_{0,n}$ when there are Ramond punctures then quite simply follow from the fact that $u_0(x)$ no longer vanishes in the $x>0$ regime.

So what is the $u_0(x)$ that we seek? The classical equation is changed from (\ref{eq:classical-one}) to:
 \begin{equation}
     u_0{\cal R}_0^2={\widetilde\Gamma}^2\ ,\quad \text{with}\quad {\cal R}_0\equiv\sum_{k=1}^\infty {\tilde t}_k u_0^k+x\ ,
     \label{eq:classical-two}
 \end{equation}
 where now the question arises as to what are the appropriate $\{{\tilde t}_k,\mu\}$. In principle, they can be different from the original $\{t_k,\mu\}$ we had when ${\widetilde\Gamma}$ is zero (as happens with {\it e.g.,} the ${\cal N}{=}2$ case). 

 \section{An Inverse Problem}
 \label{sec:inverse-problem}
 This is in fact a very clean inverse problem. Using the new $u_0(x)$ and $\mu$ in the general formulae should reproduce the formula we already have deduced in (\ref{eq:W0n2-special})--(\ref{eq:W1n4-special}), so it should provide more than enough ``data'' to constrain the result. In fact, this can be regarded as a general technique for determining the $u(x)$ associated with a model if only correlators $W_{g,n}$ are known. The general approach would be to write  expansions:
  \begin{widetext}
 \begin{eqnarray}
 &&\mu=\mu^{(0)}+\mu^{(1)}{\widetilde\Gamma}^2+\mu^{(2)}{\widetilde\Gamma}^4+\cdots\qquad
     {\tilde t}_k=t^{(0)}_k+t_k^{(1)}{\widetilde\Gamma}^2+t_k^{(2)}{\widetilde\Gamma}^4+\cdots\ ,
 \label{eq:mu-t-expansion}
 \end{eqnarray}
 and solving (\ref{eq:classical-two}) iteratively for $u_0(x)$ gives:

 \begin{equation}
\begin{aligned}
 &u_0(x)
={}
\frac{\widetilde\Gamma^2}{x^2}
-2{\tilde t}_1\frac{\widetilde\Gamma^4}{x^5}
+\widetilde\Gamma^6
\left[
-\frac{2{\tilde t}_2}{x^7}
+\frac{7{\tilde t}_1^2}{x^8}
\right]
%\\
%&
+\widetilde\Gamma^8
\left[
-\frac{2{\tilde t}_3}{x^9}
+\frac{18{\tilde t}_1{\tilde t}_2}{x^{10}}
-\frac{30{\tilde t}_1^3}{x^{11}}
\right]
\\
&\hskip8cm
+\widetilde\Gamma^{10}
\left[
-\frac{2{\tilde t}_4}{x^{11}}
+\frac{22{\tilde t}_1{\tilde t}_3+11{\tilde t}_2^2}{x^{12}}
-\frac{132{\tilde t}_1^2{\tilde t}_2}{x^{13}}
+\frac{143{\tilde t}_1^4}{x^{14}}
\right]+ \cdots\ ,
\label{eq:u0-expansion}
\end{aligned}
\end{equation}
from which derivatives can be computed and inserted into the general formulae of ref.~\cite{Johnson:2026twg}, examples of which are:
\begin{eqnarray}
    \label{eq:W03-z}
  &&  W_{0,3}=-\frac12\frac{u_0^\prime(\mu)}{z_1^2z_2^2z_3^2}\ .
\\
\label{eq:W04-z}
&&W_{0,4}(z_1,z_2,z_3,z_4)
=
\frac{1}{z_1^2z_2^2z_3^2z_4^2}
\left[
A_{04}+B_{04}
\sum_{i=1}^4 \frac{1}{{\hat z}_i^2}
\right]
=
\frac{1}{z_1^2z_2^2z_3^2z_4^2}
\left[-
\frac{u_0''(\mu)}{2}
+\frac{3\bigl(u_0'(\mu)\bigr)^2}{4}
\sum_{i=1}^4 \frac{1}{{\hat z}_i^2}
\right]
\\
\label{eq:W11-z}
&& W_{1,1}(z)
    %\nonumber\\
    =\frac{A_{11}}{z^2}+\frac{B_{11}}{z^4}
    =\frac{1}{24}\frac{u_0^{\prime\prime}(\mu)}{u_0^\prime(\mu)}\frac{1}{z^2}-\frac{u_0^\prime(\mu)}{16}  \frac{1}{z^4}\ ,
\\
\label{eq:W12-z}
&&W_{1,2}(z_1,z_2)=
\frac{1}{z_1^2 z_2^2}
\left[
A_{12}
+ B_{12}\sum_{i=1}^2 \frac{1}{{\hat z}_i^2}
+ C_{12}\sum_{i=1}^2 \frac{1}{{\hat z}_i^4}
+ D_{12}\frac{1}{z_1^2 z_2^2}
\right]
\nonumber
\\
&&\hskip-1cm \text{with}\quad
A_{12}= \frac{u_0^{(3)}(\mu)}{24\,u_0'(\mu)}-\frac{u_0''(\mu)^2}{24\,u_0'(\mu)^2}\ ,
\quad
B_{12}=-\frac{u_0''(\mu)}{8}\ ,
\quad
C_{12}=\frac{5}{32}u_0'(\mu)^2\ ,
\quad
D_{12}=\frac{3}{32}u_0'(\mu)^2\ ,
\nonumber
\\
\label{eq:W21}
\end{eqnarray}
and also:
\begin{eqnarray}
&&W_{2,1}(\mu;z)
=
\frac{A_{21}}{z^2}
+
\frac{B_{21}}{z^4}
+
\frac{C_{21}}{z^6}
+
\frac{D_{21}}{z^8}
+
\frac{E_{21}}{z^{10}} .
\\
&&\hskip-1cm \text{with}\quad
A_{21}
={}
\frac{1}{45}\,
\frac{u_0''(\mu)^4}{u_0'(\mu)^5}
-\frac{37}{960}\,
\frac{u_0''(\mu)^2u_0'''(\mu)}{u_0'(\mu)^4}
+\frac{7}{960}\,
\frac{u_0'''(\mu)^2}{u_0'(\mu)^3}
%\\
%&
+\frac{31}{2880}\,
\frac{u_0''(\mu)u_0^{(4)}(\mu)}{u_0'(\mu)^3}
-\frac{1}{576}\,
\frac{u_0^{(5)}(\mu)}{u_0'(\mu)^2} .
\nonumber
\\
%\begin{equation}
&& B_{21}
=
\frac{17}{1920}\,
\frac{u_0''(\mu)^3}{u_0'(\mu)^3}
-\frac{17}{960}\,
\frac{u_0''(\mu)u_0'''(\mu)}{u_0'(\mu)^2}
+\frac{1}{128}\,
\frac{u_0^{(4)}(\mu)}{u_0'(\mu)} .
\nonumber
%\end{equation}
%\begin{equation}
\\
&& C_{21}
=
-\frac{29}{768}\,u_0'''(\mu)
+
\frac{1}{256}\,
\frac{u_0''(\mu)^2}{u_0'(\mu)}\  ,\quad
D_{21}
=
\frac{203}{1536}\,u_0'(\mu)u_0''(\mu)\ ,\quad  
E_{21}
=
-\frac{105}{1024}\,u_0'(\mu)^3 .\nonumber
\end{eqnarray}

\end{widetext}
(Many others can be found in ref.~\cite{Johnson:2026twg}, as well as the simple method for constructing them. See also refs.~\cite{Johnson:2024bue,Ahmed:2025lxe}.)

In fact, as a test of this, it is instructive to take a few cases of the known $W_{g,n}$ for the  ${\cal N}{=}2$ case and use them to deduce a few orders of the expansion of the~${\tilde t}_k$.  Typically those are written in terms of the threshold energy $E_0$, which is related to~$\widetilde{\Gamma}$ for that model by $\widetilde{\Gamma}{=}\frac{1}{4\pi^2}\sin(2\pi\sqrt{E_0})$. This can be used to rewrite the $\tilde t_k$ as an expansion in powers of~${\widetilde\Gamma}$, to match the approach here. Then,  just studying $W_{1,1}$ and $W_{2,1}$ rapidly yields ${\mu }{=}\frac{1}{2\pi}{-}\frac{\pi}{3}E_0{+}\cdots$,  ${\tilde t}_1{=}\frac{\pi}{6}{-}\frac{\pi^3}{18}E_0{+}\cdots$ and ${\tilde t}_2{=}\frac{\pi^3}{40}{-}\frac{\pi^5}{120}E_0{+}\cdots$
which are the first few terms coming from expanding the known exact result
${t}_k{=}\frac{\pi^{k-1}J_k(2\pi\sqrt{E_0})}{2(2k+1)k!E_0^{k/2}}\ .$ So the method works.

Turning to the ${\cal N}=1$ case in hand, it is instructive to start with the general formulae given for $W_{1,1}$ and $W_{0,4}$ given  in equation~(\ref{eq:W11-z}) and~(\ref{eq:W04-z}). They both take as input the  first and second derivatives of $u_0$ which are readily computed from~(\ref{eq:u0-expansion}), and  evaluated at $\mu$, which is given in~(\ref{eq:mu-t-expansion}), where $\mu^{(0)}$ is the leading value, $1$. The result is, to order 4:
\begin{eqnarray}
    &&u_0(\mu)= \G^2 - (2\pi^2+2\mu^{(1)})\G^4+\cdots\label{eq:u0-Gamma-expansion}\\
    &&u_0^\prime(\mu) =  -2\G^2 + (10\pi^2+6\mu^{(1)})\G^4+\cdots\\
    &&u_0^{\prime\prime}(\mu) =  6\G^2 - (60\pi^2+24\mu^{(1)})\G^4+\cdots\ ,
\end{eqnarray}
and it is good to record the ratio \begin{equation}
    \frac{u_0''(\mu)}{u_0(\mu)}=-3+(15\pi^2+3\mu^{(1)})\G^2+\cdots
\end{equation}
So at zeroth order in $\G$ we recover the known purely NS result $W^{(0)}_{1,1}=-\frac{1}{8z^2}$, and at order two we get:
\begin{equation}
    W^{(2)}_{1,1} = \left(\frac{5\pi^2+\mu^{(1)}}{8}\right)\frac{1}{z^2}+\frac{1}{8}\frac{1}{z^4}\ .
\end{equation}
Comparing to the ``data'' (\ref{eq:W1n2-special}) we have 
\begin{equation}
  W_{1,1}^{(2),\mathrm{data}}
  =
  \frac{5\pi^2}{8}\frac{1}{\hat z^2}
  +\frac{5}{16}\frac{1}{\hat z^4}
  .
\end{equation}
and so if $\mu^{(1)}$ vanishes we match the first term, but the second is puzzling. The same thing happens for some of the $W_{0,4}$ results. That $W^{(0)}_{0,4}=0$ (as it should) follows since the coefficients are both at order $\G^2$ and $\G^4$. Testing second order gives
\begin{equation}
    W_{0,4}^{(2)}=-\frac12\frac{u_0^{\prime\prime}(\mu)}{z_1^2z_2^2z_3^2z_4^2}=-\frac{3}{z_1^2z_2^2z_3^2z_4^2}\ ,
\end{equation}
which matches the prediction in equation~(\ref{eq:W0n2-special}), but order~$\G^4$ yields:
\begin{equation}
  W_{0,4}^{(4)}
  =\frac{1}{z_1^2z_2^2z_3^2z_4^2}
  \left[
  12\mu^{(1)}+30\pi^2
  +3\sum_{i=1}^4\frac{1}{z_i^2}
  \right]\ .
\end{equation}
However, the  data we are matching to in~(\ref{eq:W0n4-special}) gives:
\begin{equation}
  W_{0,4}^{(4),\mathrm{data}}
  =\frac{1}{{\hat z}_1^2{\hat z}_2^2{\hat z}_3^2{\hat z}_4^2}
  \left[
  30\pi^2
  +\frac{15}{2}\sum_{i=1}^4\frac{1}{{\hat z}_i^2}
  \right]\ .
\end{equation}
so again setting $\mu^{(1)}=0$ yields the correct lowest order term, but coefficients of higher inverse powers of $z_i$ do not turn out well. There seems to be a puzzle, and it is reflected in similar computations with other examples.

The fact that the discrepancy is a factor of $\frac52$ in each of these two cases is a clue. Put differently we need another~$\frac32$ amounts of each~$\frac{1}{z_i^4}$ term. This simply comes from realizing that what is meant by the local coordinates $z_i$ {\it vs.} $\hat z_i$ is not the same in each case. For our data that came from a background with $u_0(\mu)=0$, we have ${\hat z}_i^2{=}{-}E_i$. The whole function $u_0(x)$ vanishes there, including all its derivatives. This is an earmark of a ``hard edge'' model and  so we can (and will) refer to~${\hat z}_i$ as ``hard edge'' coordinates.  Meanwhile for the new background for which we are using the universal formulae, $u_0(\mu)=E_0\neq0$, a threshold energy, and moreover the function has non-zero derivatives there. This is more akin to ``soft edge'' behaviour and so henceforth we will refer to the $z_i$ as ``soft edge'' coordinates. We have:
\begin{equation}
z_i^2{=}u_0(\mu)-E_i{=}E_0+{\hat z}_i^2\ . 
\label{eq:local-coordinate-change}
\end{equation}
We must convert between these two sets of local coordinates, and there will be potential corrections since the presence of $u_0(\mu)$ in the translation will introduce new powers of $\G^2$ (see equation~(\ref{eq:u0-Gamma-expansion})). The bottom line is that the $W_{g,n}$ being discussed come from multiplying by either Jacobian factors $\prod_i (-2z_i)$ or $\prod_i (-2{\hat z}_i)$. So to convert from one to the other one must strip off one factor and replace it by the other. To go from the general formulae $W_{g,n}(\{z_i\})$ (depending on $u_0$ and its derivatives at $\mu$) to our ``data'' $W_{g,n}\{\hat z_i\}$ we have to multiply for the $i$th leg a factor $\frac{{\hat z}_i}{z_i}$. Additionally, a term $\frac{1}{z_i^{2m}}$ becomes $\frac{1}{{\hat z}_i^{2m}}\left(1+\frac{u_0(\mu)}{{\hat z}_i^2}\right)^{-m}$. So overall, every occurrence $\frac{1}{z_i^{2m}}$ in the general non-zero $u_0$ formulae for $W_{g,n}$  must be replaced according to:
\begin{equation}
    \frac{1}{z_i^{2m}}\to \frac{1}{{\hat z}_i^{2m}}\left(1+\frac{u_0(\mu)}{{\hat z}_i^2}\right)^{-m-\frac12}\ .
\end{equation}
Let us test this. Now we see that the successful $W^{(0)}_{1,1}$ at zeroth order needs to be converted to ${\hat z}_i$ coordinate, and there is a factor $\left(1+\frac{\G^2}{{\hat z}^2}+\cdots\right)^{-\frac32}=1-\frac32\frac{\G^2}{{\hat z}^2}+\cdots$ which sends $\frac{3}{16}\frac{1}{{\hat z}^4}$ up to contribute to $W^{(2)}_{1,1}$, correcting the $\frac{1}{8}\frac{1}{{\hat z}^4}$ to $\frac{5}{16}\frac{1}{{\hat z}^4}$, {\it precisely}
what is needed for a successful match! Similarly the correction coming from converting $W^{(2)}_{0,4}$ adds a $\frac{9}{2}\sum_{i=1}^4\frac{1}{{\hat z}_i^2}$ inside the bracket in $W^{(4)}_{0,4}$, giving a total of $\frac{15}{2}$, exactly what is needed.

Much the same thing happens at this order for all the tests, and going to higher genus explores corrections to~$t_1$ and eventually the higher $t_k$. In fact, the first places the first correction to $t_1$ potentially appears are $W^{(4)}_{1,1}$ and $W^{(4)}_{1,2}$, and together they determine that $t_1^{(1)}=0$ as well as $\mu^{(2)}=0$.

At this point the only pattern that presents itself (so far) is in fact the simplest one: {\it There are no corrections to~$\mu$ or~$t_k$.} We  will assume this henceforth (since no tests contradicted this conclusion).  There will be several opportunities for this to be tested further later, and it will turn out to be correct. 

There's a certain simplicity to the $t_k$ being unchanged, and (in retrospect) perhaps it makes sense. In the BPS case of ref.~\cite{Johnson:2023ofr}, turning on $\Gamma$ backreacts on the background by changing the $t_k$. By contrast, in this case, the~$t_k$ being unchanged means that turning on $\Gamma$  leaves the background alone while adding in the new type of (R) boundary. These boundaries (really, R punctures) are  new  {\it probes} of the background. 

Finally then, the full statement of the new solution to the string equation can be made, defining the direct  matrix model description that includes both NS and R boundaries on the same footing. We will do that next.
\begin{figure}[t]
    \centering
    \includegraphics[width=0.99\linewidth]{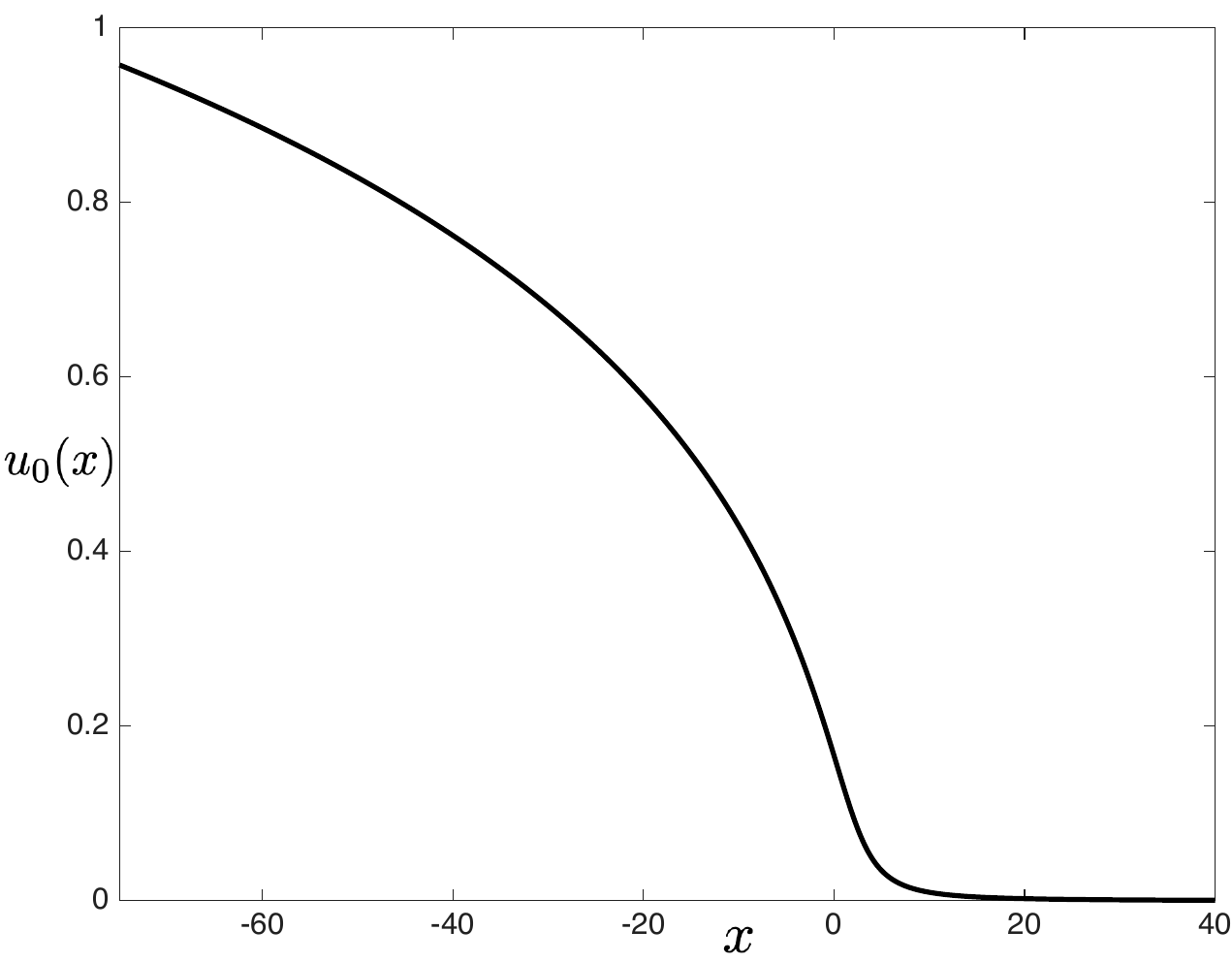}
    \caption{A plot of $u_0(x)$ from equation~(\ref{eq:compact-string-equation}), for $\G{=}1$. Now both NS boundaries and R punctures  are present.}
    \label{fig:rho-functions-with-Gamma}
\end{figure}

\section{Both Neveu-Schwarz boundaries and Ramond Punctures}
\label{sec:Ramond-from-Random}
The $\widetilde\Gamma$-deformation we seek is now easy to state: The  leading solution for $u_0(x)$ with $\G$ non-zero solves:
\begin{equation}
  \label{eq:compact-string-equation}
u_0\left(I_0\left(2\pi\sqrt{u_0}\right)-1+x\right)^2={\widetilde\Gamma}^2\ .
\end{equation}
A plot of it is given in  figure~\ref{fig:rho-functions-with-Gamma} for $\G{=}1$. Compare with figure~\ref{fig:u-functions-no-Gamma} where $\G{=}0$.
Meanwhile, $\rho_0(E)$ satisfies:
\begin{equation}
\label{eq:leading-representation-2}
    \rho_0(E)=\frac{1}{2\pi\hbar}\int_{-\infty}^1 \frac{\Theta(E-u_0(x))}{\sqrt{E-u_0(x)}} dx\ .
\end{equation}
Alternatively we can work with $u_0$ as a coordinate, giving after computing the Jacobian:
\begin{equation}
\label{eq:nice-form-u0}
    \rho_0(E) = \frac{1}{2\pi \hbar} \int_{E_0}^E \frac{f_0(u_0) du_0}{\sqrt{E-u_0}}  + \frac{\widetilde\Gamma}{4\pi \hbar} \int_{E_0}^E \frac{du_0}{u_0^{3/2} \sqrt{E-u_0}}\ , 
\end{equation}
where:
\begin{equation}
    f_0 = \frac{d}{du_0}I_0\left(2\pi\sqrt{u_0}\right)=\frac{\pi}{\sqrt{u_0}}I_1\left(2\pi\sqrt{u_0}\right)\ ,
\end{equation}
and $E_0\equiv u_0(\mu=1)$. This threshold energy  is defined by putting $u_0=E_0$ and $x=1$ into~(\ref{eq:compact-string-equation}), giving:
\begin{equation}
\G=\sqrt{E_0}\,I_0(2\pi\sqrt{E_0})\ .
\label{eq:endpoint-equation}
\end{equation}
\begin{widetext}
We then can write the final form for the spectral density as:
\begin{equation}
\rho_0(E)=
\frac{1}{2\pi\hbar}
\left[
\int_{E_0}^{E}
\frac{\pi I_1(2\pi\sqrt {u_0})}{\sqrt {u_0}\sqrt{E-u_0}}\,du_0
+
I_0(2\pi\sqrt{E_0})\frac{\sqrt{E-E_0}}{E}
\right]\ ,\qquad E>E_0\ ,
\label{eq:exact-density-i0}
\end{equation}
where the integral form seems to be the most compact presentation of the result. 
Just as a check, sending $\G\to0$ sends $E_0\to0$ while sending $\G/\sqrt{E_0}\to1$. With lower limit at $0$, inside the bracket the integral  gives $(\cosh(2\pi\sqrt{E})-1)/\sqrt{E}$
while the second term  gives $1/\sqrt{E}$ and the standard known density~(\ref{eq:N-1-density}) is recovered.

Note  that  equation~(\ref{eq:endpoint-equation}) can be inverted to write $E_0$ in terms of $\G^2$:
\begin{align}
E_0(\G)
={}&\G^2-2\pi^2\G^4+\frac{13\pi^4}{2}\G^6
-\frac{230\pi^6}{9}\G^8
+\cdots
%\frac{32053\pi^8}{288}\G^{10}
%-\frac{103341\pi^{10}}{200}\G^{12}
%+O(\G^{14})
\ ,
\label{eq:e0-expansion}
\end{align}
which will be useful later.
Using this, the whole result for the density can be expanded in powers of $\G^2$:
\begin{align}
\rho_0(E)
=\frac{1}{2\pi\hbar}\Bigg[{}
&\frac{\cosh(2\pi\sqrt E)}{\sqrt E}
-\G^2\frac{1}{2E^{3/2}}
+\G^4\frac{2\pi^2E-1}{8E^{5/2}}
-\G^6\frac{22\pi^4E^2-12\pi^2E+3}{48E^{7/2}}
\nonumber\\
&+\G^8\frac{1444\pi^6E^3-810\pi^4E^2+270\pi^2E-45}{1152E^{9/2}}
+\cdots
\Bigg]\ ,
\label{eq:density-r-expansion}
\end{align}
    but this is to be treated with care. It is a small $\widetilde\Gamma$ expansion about the usual ${\cal N}{=}1$ density on the whole positive $E$ line. The full density is defined for generic $\widetilde{\Gamma}$ and runs from finite $E_0{>}0$.
   % \vfill\eject

\end{widetext}

%\newpage

\section{The Spectral Curve, General Volumes, and Topological Recursion}
\label{sec:spectral-curve}
\subsection{The Spectral Curve incorporating Ramond punctures - Soft Edge Coordinates}
\label{sec:spectral-curve-soft}

So the leading solution for $u_0(x)$ with $\widetilde\Gamma$ turned on is simply given by~(\ref{eq:compact-string-equation}). Any  $W_{g,n}^{(2m)}$ can be written using the approach of the previous section, by using the general formulae of ref.~\cite{Johnson:2026twg}. It is instructive to recast this all in terms of the language of spectral curves. This is easy to state. Using the variable $z^2=E_0-E$, we simply define it as:
\begin{equation}
 X(z)=
 z^2-E_0\ ,
 \quad
 Y(z)=-\pi\hbar i\,\rho_0(E_0-z^2)\ ,
 \label{eq:Ramond-curve}
\end{equation}
where $\rho_0(E)$ is given in equation~(\ref{eq:exact-density-i0}). For all intents and purposes, our task is complete at this point since the $W^{(2m)}_{g,n}(\{ z_i\})$ that follow from this curve are those obtained by using the general formulae of ref.~\cite{Johnson:2026twg}, and the conversion to ``hard-edge'' coordinates~$\hat{z}_i$ give the quantities $W^{(2m)}_{g,n}(\{ \hat z_i\})$ with Ramond boundary insertions in the form we've been deducing them. 

Nevertheless, someone who prefers the topological recursion~\cite{Chekhov:2006vd,Eynard:2007kz} approach can verify this by first defining the standard initial data for the method:
\begin{equation}
 \omega_{0,1}(z)=Y(z)dX(z),
 \qquad
 \omega_{0,2}(z_1,z_2)=B(z_1,z_2),
 \label{eq:tr-initial-data}
\end{equation}
where $B(z_1,z_2)$ is the standard (Cauchy) %genus-zero 
Bergman kernel on the sphere:
\begin{equation}
B(z_1,z_2)=\frac{dz_1\,dz_2}{(z_1-z_2)^2}\ ,
 \label{eq:bergman-kernel}
\end{equation} 
and 
%\begin{equation}
$dX(z)=2z\,dz$,
%\end{equation} 
where there is the local involution 
\begin{equation}
    \sigma(z)=-z
    \label{eq:local-involution}
\end{equation}
with a branch point at $z{=}0$.
This is all then a straightforward case for crunching the  recursion, which
(for $2g{-}2{+}n{>}0$) is:
\begin{widetext}

\begin{align}
 \omega_{g,n}(z_1,z_S)
 ={}&\operatorname*{Res}_{z=0} K(z_1,z)
 \Bigg[
 \omega_{g-1,n+1}(z,\sigma(z),z_S)
 %\nonumber\\
% &\hspace{3.4cm}
+\sum_{\substack{g_1+g_2=g\\ I\sqcup J=S}}^{\prime}
 \omega_{g_1,|I|+1}(z,z_I)
 \omega_{g_2,|J|+1}(\sigma(z),z_J)
 \Bigg].
 \label{eq:tr-recursion}
\end{align}
    
\end{widetext}
 and the prime on the sum means that the unstable terms involving $\omega_{0,1}$ are omitted.  The recursion kernel is: \begin{equation}
 K(z_1,z)=
 -\frac{\displaystyle\int_{\sigma(z)}^{z}B(z_1,\zeta)}
 {2\bigl(Y(z)-Y(\sigma(z))\bigr)dX(z)}\ .
 \label{eq:kernel-def}
\end{equation}
Since $Y(\sigma(z))=-Y(z)$ (further discussed  below) and:
\begin{equation}
 \int_{-z}^{z}B(z_1,\zeta)
 =\frac{2z\,dz_1}{z_1^2-z^2}\ ,
 \label{eq:kernel-pieces}
\end{equation}
the recursion kernel becomes:
\begin{equation}
 K(z_1,z)=-\frac{dz_1}{4Y(z)(z_1^2-z^2)\,dz}\ .
 \label{eq:local-kernel}
\end{equation}
Once the $\omega_{g,n}$ are obtained, the $W_{g,n}$s are defined by:
\begin{equation}
 \omega_{g,n}(z_1,\ldots,z_n)
 =W_{g,n}(z_1,\ldots,z_n)\prod_{i=1}^{n}dz_i.
 \label{eq:omega-to-w}
\end{equation}
The novelty here is that $Y(z)$ is defined through an integral (see~(\ref{eq:exact-density-i0}) with $E$ replaced by $E_0-z^2$), but the standard treatment applies, since all that matters is the local behaviour near $z{=}0$, which is readily obtained by expanding  the integral near there.

The overall point is that the topological recursion procedure {\it must} yield the same results as just taking the $u_0(x)$ from which the spectral curve is derived and using it in the general formulae for $W_{g,n}$ derived in ref.~\cite{Johnson:2026twg}, which are built from combinations of the $x$-derivatives of~$u_0$.
Indeed, let us pause to show that this is a quite general correspondence, true for any $Y(z)$ obtained from a spectral density that has the integral representation:
\begin{equation}
\label{eq:nice-form-2}
    \rho_0(E) = \frac{1}{2\pi \hbar} \int_{E_0}^E \frac{f(u_0) du_0}{\sqrt{E-u_0}} \ , \quad f(u_0)\equiv-\frac{dx}{du_0}\ .
\end{equation}
This deserves a subsection of its own, which follows.

\subsection{Ref.~\cite{Johnson:2026twg}'s General $W_{g,n}$ Formulae and Topological Recursion}
\label{sec:TR-and-the-general-formulae}
\noindent The  curve ($X(z),Y(z)$)  that follows from continuing the general integral representation~(\ref{eq:nice-form-2}) is:
 \begin{equation}
 X(z)=
 z^2-E_0\ ,
 \quad
 Y(z)=\frac{1}{2}\int_0^{z^2}\frac{f(E_0-v)\,dv}{\sqrt{z^2-v}}\ ,
 \label{eq:y-local-integral}
\end{equation}
where we used the substitution $u_0{=}E_0{-}v$.  Taylor expanding: \begin{equation}
    f(E_0-v)=\sum_{m=0}^\infty\frac{(-1)^m}{m!}\left.\frac{d^m\!f}{du_0^m}\right|_{E_0} v^m\ ,
\end{equation} the integral boils down to a sum of integrals that can be readily evaluated:
\begin{eqnarray}
    &&\int_0^{z^2}\!\!\frac{v^m}{\sqrt{z^2-v}}dv =\\
    &&\hskip0.7cm=z^{2m+1}\!\!\int_0^1\!\!\frac{w^m}{\sqrt{1-w}}dw=\boldsymbol{\beta}\left(m+1,\frac12\right)z^{2m+1}\ ,
    \nonumber
\end{eqnarray} where  here $\boldsymbol{\beta}(a,b)\equiv\Gamma(a)\Gamma(b)/\Gamma(a+b)$ is the Euler-Beta function, and the substitution $v=z^2w$ was used. Here, $\boldsymbol{\beta}(1,\frac12){=}2$, $\boldsymbol{\beta}(2,\frac12){=}\frac43$, $\boldsymbol{\beta}(3,\frac12){=}\frac{16}{15}$, and so on.
So in fact:
\begin{equation}
 Y(z)=\frac{1}{2}\sum_{m=0}^{\infty}
 \frac{(-1)^m}{m!}
 \left.\frac{d^m\!f}{du_0^m}\right|_{E_0} 
 \boldsymbol{\beta}\!\left(m+1,\frac{1}{2}\right)z^{2m+1}\ .
 \label{eq:y-series-general}
\end{equation}
Now since $f(u_0){\equiv}{-}{dx}/{du_0}{=}{-}{1}/{u_0^\prime}$ we can write:
\begin{equation}
    \frac{df}{du_0}=\frac{u_0^{\prime\prime}}{(u_0^\prime)^3}\ ,\quad\text{and}\quad 
    \frac{d^2f}{du_0^2}=\frac{u_0^\prime u_0^{\prime\prime\prime}-3(u_0^{\prime\prime})^2}{(u_0^\prime)^5}\ ,
\end{equation}
and so on.
So finally we have:
\begin{equation}
 Y(z)=-\frac{1}{u_0^\prime}z-\frac23\frac{u_0^{\prime\prime}}{(u_0^\prime)^3} z^3+\frac{4}{15}\frac{u_0^\prime u_0^{\prime\prime\prime}-3(u_0^{\prime\prime})^2}{(u_0^\prime)^5} z^5+O(z^7)\ ,
 \label{eq:Yz-expansion}
\end{equation}
and the various combinations of $x$-derivatives of $u_0$ are evaluated at $x{=}\mu$.

In the topological recursion procedure, the $\omega_{g,n}$ will result from extracting the pole behaviour resulting from the kernel multiplying other $\omega_{g',n'}$. Since the kernel has~$Y(z)$ in the denominator (see equation~(\ref{eq:local-kernel})) combinations of the $u_0$ derivatives in the expansion~(\ref{eq:Yz-expansion}) will find their way into expressions for the $\omega_{g,n}$, and hence the $W_{g,n}$. This is, quite generally,  topological recursion's way of reproducing the general formulae of ref.~\cite{Johnson:2026twg}!

Let's check. The simplest case is to derive $W_{0,3}$. The recursion gives:

\begin{widetext}
\begin{eqnarray}
 \omega_{0,3}(z_1,z_2,z_3)
 &=&\operatorname*{Res}_{z=0}K(z_1,z)
 \Bigl[
 \omega_{0,2}(z,z_2)\omega_{0,2}(\sigma(z),z_3)
 %\nonumber\\
 %&\hspace{2.5cm}
 +\omega_{0,2}(z,z_3)\omega_{0,2}(\sigma(z),z_2)
 \Bigr]
 \label{eq:omega03-recursion}
\nonumber\\
% \omega_{0,3}(z_1,z_2,z_3)
 &=&\operatorname*{Res}_{z=0}
 \left[
 \frac{dz\,dz_1dz_2dz_3}
 {4Y(z)(z_1^2-z^2)(z-z_2)^2(z+z_3)^2}
 +\frac{dz\,dz_1dz_2dz_3}
 {4Y(z)(z_1^2-z^2)(z-z_3)^2(z+z_2)^2}
 \right]\ .
 \label{eq:omega03-integrand}
\end{eqnarray}
\end{widetext}
But here it is clear that only  $Y(z){=}-z/u_0^\prime+O(z^3)$ contributes to the residue and the two terms give equal contributions, yielding:
\begin{equation}
 \omega_{0,3}(z_1,z_2,z_3)
 =-\frac{u_0^\prime}{2}\frac{dz_1dz_2dz_3}{z_1^2z_2^2z_3^2}\ ,
 \label{eq:omega03-a}
\end{equation}
which exactly reproduces the general equation~(\ref{eq:W03-z}).
Meanwhile  $W_{1,1}(z_1)$ follows even more simply because:
\begin{equation}
 \omega_{1,1}(z_1)=\operatorname*{Res}_{z=0}K(z_1,z)\omega_{0,2}(z,\sigma(z))\ .
 \label{eq:omega11-recursion}
\end{equation}
Since $\omega_{0,2}(z,\sigma(z))=B(z,\sigma(z))$ and 
\begin{equation}
 B(z,\sigma(z))=B(z,-z)=\frac{dz\,d(-z)}{(z+z)^2}
 =-\frac{dz^2}{4z^2}\ ,
 \label{eq:B-z-minus-z}
\end{equation}
we get:
\begin{equation}
 \omega_{1,1}(z_1)
 =\operatorname*{Res}_{z=0}\left[\frac{dz\,dz_1}{16z^2Y(z)(z_1^2-z^2)}\right]\ .
 \label{eq:omega11-integrand}
\end{equation}
Now because of the extra power of $z$ in the denominator we must expand, using:
\begin{eqnarray}
 \frac{1}{Y(z)}&=&-\frac{u_0^\prime}{z}\left(1-\frac23\frac{u_0^{\prime\prime}}{(u_0^\prime)^2}z^2+O(z^4)\right),
 \nonumber \\
 \frac{1}{z_1^2-z^2}&=&\frac{1}{z_1^2}\left(1+\frac{z^2}{z_1^2}+O(z^4)\right)\ .
 \label{eq:omega11-expansions}
\end{eqnarray}
The residue that results from gathering all the pieces together is:
\begin{equation}
 \omega_{1,1}(z_1)=
 \left[\frac{1}{24}\frac{u_0''(\mu)}{u_0'(\mu)}\frac{1}{z_1^2}
 -\frac{u_0'(\mu)}{16}\frac{1}{z_1^4} \right]dz_1\ ,
 \label{eq:omega11}
\end{equation} which precisely yields equation~(\ref{eq:W11-z}) of the general formulae. It is straightforward to continue this exercise, showing how higher terms in the expansion of $Y(z)$ combine to yield, through the recursion, the formulae of ref.~\cite{Johnson:2026twg}, as they must. (The cases of $W_{0,4}$ and $W_{1,2}$ are included in  Appendix~\ref{app:W-04-and-W-12}.) Of course, it bears repeating here (as emphasized there) that topological recursion takes the scenic route to arriving at the formula for $W_{g,n}$. The fact that $W_{g,n}$ follows directly from $W_{g,n-1}$ by use of ref.~\cite{Johnson:2026twg}'s simple operator $\delta_{E_n}^{(u_0)}$ (derived from the KdV flows) that acts simply on the $u_0$-dependence, is invisible in the standard topological recursion approach. See more discussion of this in   Section (\ref{sec:discussion}).

Returning to our matrix model including R-punctures,  the natural spectral curve is indeed that given in equation~(\ref{eq:Ramond-curve}), and either using topological recursion, or directly the general formulae in terms of $u_0$ and its derivatives, the $W^{(2m)}_{g,n}(\{z_i\})$ can be readily written. These must be converted to the ``hard edge'' coordinates $\hat z_i$ to get $W^{(2m)}_{g,n}(\{\hat z_i\})$, by changing variables and remembering the leg factors, as explained in Section~\ref{sec:inverse-problem}.

It might be natural to ask if one can express the spectral curve (and the resulting topological recursion computation) {\it directly} in the natural ``hard edge'' coordinates, avoiding the conversion at the end. The answer is yes, but the principle of conservation of difficulty applies, of course. The leg-factor conversion procedure will be replaced by something else. We turn to that next.

\subsection{The Spectral Curve incorporating Ramond punctures - Hard Edge Coordinates}
\label{sec:spectral-curve-hard}
To define the  spectral curve  directly in the hard edge coordinate $\hat z$, the first step is to rewrite the spectral curve by defining:
\begin{equation}
    X(\hat z)=\hat z^2\ ,\quad 
    Y(\hat z)=-\pi\hbar i\,\rho_0(-\hat{z}^2)\ ,
    \label{eq:new-Ramond-curve}
\end{equation}
where again $\rho_0(E)$ is given in equation~(\ref{eq:exact-density-i0}). The hard edge disc data are now in $\omega_{0,1}{=}Y(\hat z)dX(\hat z)$.   Notice that if we construct it and use $-E=\hat z^2$ in expansion~(\ref{eq:density-r-expansion}), the Ramond part gives precisely the $W_{0,1}^{\rm R}$ that we extracted in~(\ref{eq:Ramond-disc-expansion})! As already mentioned (below that equation) ref.~\cite{norbury2024superweilpeterssonmeasuresmoduli} has an equation yielding the same quantity, but here, we now understand from the matrix model approach a different closed form equation  for it, (\ref{eq:exact-density-i0}), that together with~(\ref{eq:endpoint-equation}) yields the  definition of the spectral curve directly in hard-edge terms.

We can now work with this new spectral curve, and~$\omega_{0,1}(\hat z)$ data, to compute {\it via} topological recursion but it is crucial to realize that this will give the {\it wrong answers} unless the Bergman kernel is also transformed away from the simple Cauchy case given in equation~(\ref{eq:bergman-kernel}). Explicitly the new hard edge kernel is:
\begin{eqnarray}
{\widehat B}({\hat z}_1,{\hat z}_2)
&=&
\frac{{\hat z}_1{\hat z}_2}{z_1z_2}
\frac{d{\hat z}_1d{\hat z}_2}{(z_1-z_2)^2}\ ,
\nonumber\\
\text{with}\quad
z_i&=&\sqrt{{\hat z}_i^{\,2}+E_0} \ ,   
\end{eqnarray}
where we used that $z'({\hat z})={{\hat z}}/{z}$. However,  this is deceptively simple in its presentation. In the active use of this new kernel, when writing everything in terms of $\hat z_i$ using the second line, non-trivial $E_0$  dependence will enter. Expanding in $E_0$ gives:
\begin{widetext}

\begin{equation}
{\widehat B}({\hat z}_1,{\hat z}_2)
=
\left[
\frac{1}{({\hat z}_1-{\hat z}_2)^2}
-\frac{E_0}{2{\hat z}_1^{\,2}{\hat z}_2^{\,2}}
+\frac{3E_0^2({\hat z}_1^{\,2}+{\hat z}_2^{\,2})}
{8{\hat z}_1^{\,4}{\hat z}_2^{\,4}}
+O(E_0^3)
\right]d{\hat z}_1d{\hat z}_2\ .
\end{equation}
Then substituting the solution's expansion of $E_0$ given in~(\ref{eq:e0-expansion})
gives:
\begin{equation}
{\widehat B}({\hat z}_1,{\hat z}_2)
=
\left[
\frac{1}{({\hat z}_1-{\hat z}_2)^2}
-\G^2\left(\frac{1}{2{\hat z}_1^{\,2}{\hat z}_2^{\,2}}\right)
+\G^4\left(
\frac{\pi^2}{{\hat z}_1^{\,2}{\hat z}_2^{\,2}}
+
\frac{3({\hat z}_1^{\,2}+{\hat z}_2^{\,2})}
{8{\hat z}_1^{\,4}{\hat z}_2^{\,4}}\right)+O(\G^6)
\right]d{\hat z}_1d{\hat z}_2\ .
\label{eq:hard-edge-Bergman}
\end{equation}
\end{widetext}
In other words, one can avoid the procedure of changing  variables  from $z_i$ to $\hat z_i$ (and leg-factors) and work directly with the spectral curve written in terms of $\hat z$, but the price paid is a more  cumbersome Bergman kernel that must be expanded to successively higher powers in $\G^2$, in order to correctly extract the  Ramond contributions. Which  of the equivalent approaches to take is a matter of personal taste.

The bottom line is that we've now defined the data that can be used, if desired, to use topological recursion procedures to compute all the $W^{(2m)}_{g,n}$ (and hence the volumes $V^{(2m)}_{g,n}$) recursively, something that has been missing from the literature to date. Given how it was derived here, it is clear that it must work, but
the proof of the pudding is to simply dive into the procedure and check some non-trivial cases. It is prudent to expand all the  hard-edge  data needed for the recursion procedure in one place. We already have (in equation~(\ref{eq:hard-edge-Bergman})) the Bergman kernel to order $\widetilde\Gamma^4$ and so to join it we have, by just substituting $E=-\hat z^2$ as before: 
\begin{eqnarray}
\hskip-0.5cm Y(\hat z)&=&Y^{(0)}(\hat z)+\widetilde\Gamma^2Y^{(2)}(\hat z)
+\widetilde\Gamma^4Y^{(4)}(\hat z)+\cdots,
\label{eq:hard-edge-Y-expansion}
\end{eqnarray}
with: 
\begin{eqnarray}
 Y^{(0)}(\hat z)&=&-\frac{\cos(2\pi \hat z)}{2\hat z}\ ,\quad
Y^{(2)}(\hat z)=-\frac{1}{4\hat z^3}\ ,\nonumber\\
 Y^{(4)}(\hat z)&=&\frac{1+2\pi^2\hat z^2}{16\hat z^5}\ .
\end{eqnarray}
The recursion kernel has an expansion we can write as:
\begin{widetext}
\begin{equation}
K(\hat z_1,\hat z)
=
K^{(0)}(\hat z_1,\hat z)
+
\widetilde\Gamma^2K^{(2)}(\hat z_1,\hat z)
+
\widetilde\Gamma^4K^{(4)}(\hat z_1,\hat z)
+O(\widetilde\Gamma^6)\ ,
\end{equation}
and using the definition~(\ref{eq:kernel-def})
we get, after some algebra:
\begin{equation}
K^{(0)}(\hat z_1,\hat z)
=
\frac{\hat z\,d\hat z_1}
{2\cos(2\pi\hat z)(\hat z_1^2-\hat z^2)d\hat z}\ ,
\quad
K^{(2)}(\hat z_1,\hat z)
=
\left[
\frac{1}{4\cos(2\pi\hat z)\hat z_1^2\hat z}
-
\frac{1}{4\cos^2(2\pi\hat z)\hat z(\hat z_1^2-\hat z^2)}
\right]\frac{d\hat z_1}{d\hat z}\ ,
\end{equation}
and a longer expression for $K^{(4)}(\hat z_1,\hat z)$ involving further inverse powers of $\cos(2\pi\hat z)$ and so forth that isn't explicitly needed here. What's really needed is the behaviour near \(\hat z=0\) and so, expanding:
\begin{align}
K^{(0)}(\hat z_1,\hat z)
&=
\left[
\frac{\hat z}{2\hat z_1^2}
+
\left(
\frac{\pi^2}{\hat z_1^2}
+
\frac{1}{2\hat z_1^4}
\right)\hat z^3
+O(\hat z^5)
\right]\frac{d\hat z_1}{d\hat z}\ ,
\label{eq:K0-expansion}
\\
K^{(2)}(\hat z_1,\hat z)
&=
\left[
-\left(
\frac{\pi^2}{2\hat z_1^2}
+
\frac{1}{4\hat z_1^4}
\right)\hat z
+O(\hat z^3)
\right]\frac{d\hat z_1}{d\hat z}\ ,
\label{eq:K2-expansion}
\\
\text{and}\quad
K^{(4)}(\hat z_1,\hat z)
&=
\left[
\left(
\frac{11\pi^4}{8\hat z_1^2}
+
\frac{3\pi^2}{4\hat z_1^4}
+
\frac{3}{16\hat z_1^6}
\right)\hat z
+O(\hat z^3)
\right]\frac{d\hat z_1}{d\hat z}\ .
\label{eq:K4-expansion}
\end{align}
Turning to the computation of $W^{(2)}_{0,3}$,  the recursion instructs us to compute:
\begin{align}
\omega_{0,3}(\hat z_1,\hat z_2,\hat z_3)
=
\operatorname*{Res}_{\hat z=0}K(\hat z_1,\hat z)
\Big[
\widehat B(\hat z,\hat z_2)\widehat B(-\hat z,\hat z_3)
+\widehat B(\hat z,\hat z_3)\widehat B(-\hat z,\hat z_2)
\Big]\ .
\label{eq:hard-edge-threepoint-recursion}
\end{align}
At order \(\widetilde\Gamma^0\), we recover the known result that $W^{(0)}_{0,3}=0$ since everything at that order in the bracket is regular while $K^{(0)}\sim \hat z$, giving zero residue. Turning to  order \(\widetilde\Gamma^2\), there are two possible contributions, one giving zero residue for the same structural reasons as before, and then two terms of the form
$K^{(0)}
\bigl[
\widehat B^{(2)}\widehat B^{(0)}
+
\widehat B^{(0)}\widehat B^{(2)}
\bigr]$.
We use:
\begin{equation}
\widehat B^{(2)}(\hat z,\hat z_i)
=
-\frac{d\hat z\,d\hat z_i}{2\hat z^2\hat z_i^2}\ ,
\qquad
\widehat B^{(2)}(-\hat z,\hat z_i)
=
\frac{d\hat z\,d\hat z_i}{2\hat z^2\hat z_i^2}\ ,    
\end{equation}
to get, at this order,
\begin{align}
&
\left[\widehat B^{(2)}(\hat z,\hat z_2)\widehat B^{(0)}(-\hat z,\hat z_3)
+
\widehat B^{(0)}(\hat z,\hat z_2)\widehat B^{(2)}(-\hat z,\hat z_3)\right]
+
[\hat z_2\leftrightarrow\hat z_3]
%\widehat B^{(2)}(\hat z,\hat z_3)\widehat B^{(0)}(-\hat z,\hat z_2)
%+
%\widehat B^{(0)}(\hat z,\hat z_3)\widehat B^{(2)}(-\hat z,\hat z_2)
%\nonumber\\
%&\hspace{2cm}
=
\frac{2\,d\hat z^2d\hat z_2d\hat z_3}
{\hat z^2\hat z_2^2\hat z_3^2}
+O(\hat z^0)\ .
\end{align}
Multiplying by the first term of $K^{(0)}(\hat z_1,\hat z)$ in (\ref{eq:K0-expansion}) gives:
\begin{equation}
\omega^{(2)}_{0,3}(\hat z_1,\hat z_2,\hat z_3)
=
\frac{d\hat z_1d\hat z_2d\hat z_3}
{\hat z_1^2\hat z_2^2\hat z_3^2}\quad\longrightarrow\quad
W^{(2)}_{0,3}(\hat z_1,\hat z_2,\hat z_3)
=
\frac{1}{\hat z_1^2\hat z_2^2\hat z_3^2}\ ,
\end{equation}
which is indeed the $n=3$ case of  the general formula~(\ref{eq:W0n2-special}).

\noindent Moving on to derive $W^{(4)}_{0,3}$, 
at order \(\widetilde\Gamma^4\), the same recursion~(\ref{eq:hard-edge-threepoint-recursion}) gives non-zero terms of the following structure:
\begin{equation}
K^{(2)}[\widehat B^{(2)}\widehat B^{(0)}+\widehat B^{(0)}\widehat B^{(2)}]
+
K^{(0)}[\widehat B^{(4)}\widehat B^{(0)}+\widehat B^{(0)}\widehat B^{(4)}+\widehat B^{(2)}\widehat B^{(2)}]\ ,
\end{equation}
with the result for the square bracket terms being, respectively:
\begin{align}
\frac{2\,d\hat z^2d\hat z_2d\hat z_3}
{\hat z^2\hat z_2^2\hat z_3^2}
+O(\hat z^0)\ ,\quad\text{and}\quad
\left[
-\frac{2}{\hat z^4\hat z_2^2\hat z_3^2}
-
\frac{1}{\hat z^2}
\left(
\frac{4\pi^2}{\hat z_2^2\hat z_3^2}
+
\frac{3}{\hat z_2^2\hat z_3^4}
+
\frac{3}{\hat z_2^4\hat z_3^2}
\right)\right]d\hat z^2d\hat z_2d\hat z_3
+O(\hat z^0) \ ,
\end{align}
Multiplying by the $K^{(2)}$ and $K^{(4)}$ in equations~(\ref{eq:K2-expansion}) and~(\ref{eq:K4-expansion}) gives six contributions that can be summed to yield:
\begin{equation}
\omega^{(4)}_{0,3}(\hat z_1,\hat z_2,\hat z_3)
=
-\frac{d\hat z_1d\hat z_2d\hat z_3}
{\hat z_1^2\hat z_2^2\hat z_3^2}
\left[
5\pi^2+\frac{3}{2}\sum_{i=1}^{3}\frac{1}{\hat z_i^2}
\right]\ ,
\end{equation}
and indeed the $W^{(4)}_{0,3}$ that is read off is precisely the $n=4$ specialization of formula~(\ref{eq:W4n-special}),
with \(t_1=\pi^2\).
\end{widetext}
Finally, computing $W^{(2)}_{1,1}$ is very straightforward. The recursion instructs us to compute:
\begin{equation}
\omega_{1,1}(\hat z_1)
=
\operatorname*{Res}_{\hat z=0}
K(\hat z_1,\hat z)
\widehat B(\hat z,-\hat z).
\end{equation}
and for this order we have the two terms,
$K^{(2)}B^{(0)}$ and $K^{(0)}B^{(2)}$, to compute. We have, as before:
\begin{equation}
\widehat B^{(0)}(\hat z,-\hat z)
=
-\frac{d\hat z^2}{4\hat z^2},
\qquad
\widehat B^{(2)}(\hat z,-\hat z)
=
\frac{d\hat z^2}{2\hat z^4}\ ,
\end{equation}
and this yields:
\begin{align}
K^{(2)}\widehat B^{(0)}
&=
\left[
\frac{\pi^2}{8\hat z_1^2}
+
\frac{1}{16\hat z_1^4}
\right]\frac{d\hat z_1\,d\hat z}{\hat z}
+O(\hat z)\,d\hat z,
\end{align}
and
\begin{align}
K^{(0)}\widehat B^{(2)}
&=
\left[
\frac{1}{4\hat z_1^2\hat z^3}
+
\left(
\frac{\pi^2}{2\hat z_1^2}
+
\frac{1}{4\hat z_1^4}
\right)\frac{1}{\hat z}\right]d\hat z_1d\hat z
+O(\hat z)d\hat z\ 
,
\end{align}
where the first term will contribute zero to the residue. Summing the non-zero residues gives:
\begin{equation}
\omega^{(2)}_{1,1}(\hat z_1)
=
\left[
\frac{5\pi^2}{8\hat z_1^2}
+
\frac{5}{16\hat z_1^4}
\right]d\hat z_1\ ,
\end{equation}
which indeed yields the result for $W^{(2)}_{1,1}(\hat z)$ that is the \(n{=}1\) specialization of general formula~(\ref{eq:W1n2-special}).
Other examples can be computed, but it is clear at this point that this way of presenting the spectral curve data will yield the desired $W^{(2m)}_{g,n}$ through topological recursion methods, as claimed.

\section{Summary}
\label{sec:discussion}
In summary, presented here is a complete random matrix model whose correlators compute the volumes $V^{(2m)}_{g,n}(\{b_i\})$ of the compactified moduli space of super-Reimann surfaces with~$n$ Neveu-Schwarz (NS) geodesic boundaries of  lengths~$\{b_i\}$ ($i{=}1,..,n$) and $2m$ Ramond (R) punctures. The entire content (not just perturbatively but non-perturbatively) of the random matrix model  is expressed through the appropriate solution of the string equation~(\ref{eq:big-string-equation}), an approach that places the two sectors on the same footing. 

Although the presentation focused on  the computation of $W^{(2m)}_{g,n}(\{\hat z_i\})$, they are related to the $V^{(2m)}_{g,n}(\{b_i\})$ by simple Laplace transform {\it via}:
\begin{equation}
    W^{(2m)}_{g,n}(\{\hat z_i\}) = \int \prod b_i db_i V^{(2m)}_{g,n}(\{b_i\}){\rm e}^{-b_i{\hat z}_i}\ ,
\end{equation}
so the volumes are understood as being defined through the discussion of $W^{(2m)}_{g,n}$s.

The most natural and swift way to compute is to hold fixed $\G\equiv\hbar\Gamma$ in the classical limit, extract the leading part of the solution, $u_0(x)$, and compute its derivatives at $x{=}1$. Then, treating the results as an expansion in~$\G^2$, the general formulae of ref.~\cite{Johnson:2026twg} allow the results to be swiftly written down for any number of boundaries and punctures, after a change of variables.

The function $u_0(x)$ can instead be used in an integral transform to write the matrix model's leading spectral density $\rho_0(E)$, from which a spectral curve was  written (filling a  gap in the literature). The methods of topological recursion can then be applied to this spectral curve, as demonstrated explicitly. 

Ref.~\cite{norbury2024superweilpeterssonmeasuresmoduli} derived a direct volume recursion  relation for the cases with R-punctures, in the manner of Mirzakhani~\cite{Mirzakhani:2006fta}  (generalizing the work of Stanford and Witten's~\cite{Stanford:2019vob}).  It is natural to conjecture that, in the spirit of ref.~\cite{Eynard:2007fi}, there is a direct Laplace transform to be done on the hard-edge spectral curve data presented in Section~\ref{sec:spectral-curve-hard} that would yield ref.~\cite{norbury2024superweilpeterssonmeasuresmoduli}'s recursion. That exploration is left  for future work.

That one  can use  topological recursion  to reconstruct the general formulae that were derived in  ref.~\cite{Johnson:2026twg}  using  the KdV (and string equation)-based approach  was   demonstrated in Section~\ref{sec:TR-and-the-general-formulae}. It should be noted that this was guaranteed to be true, although precisely how it worked was nice to see. It is especially key to note that being able to derive them using topological recursion does not mean that the simple relations between different $W_{g,n}$ observed in ref.~\cite{Johnson:2026twg} can be seen in that framework.

Moreover, it is the string equation approach that naturally gives swift derivation of the  the striking {\it closed-form} (for all $n$) expressions that follow from the KdV-derived boundary adding operator. As observed in ref.~\cite{Johnson:2026twg}, for the supersymmetric case, the fact that the solutions are all  series in inverse powers of $x$ allows for swift deduction of these closed forms across $n$, something that is not obviously available in the topological recursion approach. Of course, now such structures have been found, it could be interesting to see if there is a way of making them visible using the tools of topological recursion. That is also an intriguing future direction.

Having the closed form formulae also revealed many striking relations between volumes in different sectors. It will certainly be interesting to understand these in a direct geometrical approach.

%\lipsum[1-6]

\begin{acknowledgments}
%\section*
CVJ   thanks  the  US Department of Energy (under grant \protect{DE-SC}~0011702)  for  support,  and  Amelia for her support and patience.    
\end{acknowledgments}

\bigskip
\appendix

\begin{widetext}
\section{Ref.~\cite{Johnson:2026twg}'s general formulae for $W_{0,4}$ and $W_{1,2}$ using topological recursion}
\label{app:W-04-and-W-12}

For what is needed for these computations using recursion~(\ref{eq:tr-recursion}), the  kernel $K(z_1,z)$ was computed as~(\ref{eq:local-kernel}), with an expansion for $Y(z)$ given in equation~(\ref{eq:Yz-expansion}). For brevity in what is to come, it is useful to define a shorthand:
\begin{equation}
Y(z)=a z+ bz^3+cz^5+\cdots\quad\text{where}\quad
%\end{equation}
%
%\begin{equation}
 a\equiv-\frac{1}{u_0^\prime}\ , \quad b\equiv -\frac23\frac{u_0^{\prime\prime}}{(u_0^\prime)^3} \ ,\quad\text{and}\quad c \equiv\frac{4}{15}\frac{u_0^\prime u_0^{\prime\prime\prime}-3(u_0^{\prime\prime})^2}{(u_0^\prime)^5} \ .
 \label{eq:Y-abc}
\end{equation}and we will need:
\begin{equation}
 \frac{1}{Y(z)}=\frac{1}{az}\left[1-\frac{b}{a}z^2+
 \left(\frac{b^2}{a^2}-\frac{c}{a}\right)z^4+O(z^6)\right].
 \label{eq:Y-inverse}
\end{equation}

\subsection{The case of $W_{0,4}$}
\label{app:W-04-and-TR}
In the sum in the recursion~(\ref{eq:tr-recursion}), let $S=\{2,3,4\}$.  For $(g,n)=(0,4)$, the right hand side only contains~$\omega_{0,3}$ and~$\omega_{0,2}=B$:
\begin{align}
 \omega_{0,4}(z_1,z_2,z_3,z_4)
 ={}&\operatorname*{Res}_{z=0}K(z_1,z)
 \sum_{k\in S}
 \Bigl[
 \omega_{0,3}(z,z_{S\setminus k})B(\sigma(z),z_k)
 \nonumber\\
 &\hspace{3.2cm}
 +B(z,z_k)\omega_{0,3}(\sigma(z),z_{S\setminus k})
 \Bigr]\ .
 \label{eq:omega04-recursion}
\end{align}
Here $z_{S\setminus k}$ denotes the two variables in $S$ different from $k$.  From equation~\eqref{eq:omega03-a},
\begin{equation}
 \omega_{0,3}(z,z_i,z_j)
 =\frac{dz\,dz_i\,dz_j}{2a\,z^2z_i^2z_j^2}\ ,
 \qquad
 \omega_{0,3}(\sigma(z),z_i,z_j)
 =-\frac{dz\,dz_i\,dz_j}{2a\,z^2z_i^2z_j^2}\ .
 \label{eq:omega03-z-sigma-for-04}
\end{equation}
The second sign follows from $d\sigma=-dz$.  Combining this with equation~\eqref{eq:B-z-minus-z}, the bracket in equation~\eqref{eq:omega04-recursion} is
\begin{equation}
 -\frac{dz^2\,dz_2dz_3dz_4}{2a}
 \sum_{k\in S}\frac{1}{z^2\prod_{i\in S\setminus k}z_i^2}
 \left[\frac{1}{(z+z_k)^2}+\frac{1}{(z-z_k)^2}\right].
 \label{eq:omega04-bracket}
\end{equation}
Therefore a typical $k$ contribution to the  residue is:
\begin{equation}
 \operatorname*{Res}_{z=0}
 \frac{dz}{8a}\,
 \frac{1}{Y(z)(z_1^2-z^2)}
 \frac{1}{z^2\prod_{i\in S\setminus k}z_i^2}
 \left[\frac{1}{(z+z_k)^2}+\frac{1}{(z-z_k)^2}\right].
 \label{eq:omega04-k-contribution}
\end{equation}
The   expansions we need are~(\ref{eq:Y-inverse}) and:
\begin{equation}
 \frac{1}{z_1^2-z^2}=\frac{1}{z_1^2}\left(1+\frac{z^2}{z_1^2}+O(z^4)\right)\ ,\quad\text{and}
 \quad
 \frac{1}{(z+z_k)^2}+\frac{1}{(z-z_k)^2}
 =\frac{2}{z_k^2}\left(1+3\frac{z^2}{z_k^2}+O(z^4)\right).
 \label{eq:W04-expansion-2}
\end{equation}
 The explicit $z^{-2}$ in equation~\eqref{eq:omega04-k-contribution} combines with $1/Y(z)$ to give $z^{-3}$, so the residue is obtained from the $z^2$ term in the remaining even expansion.  The $k$ term is therefore
\begin{equation}
 \frac{1}{4a^2}\,
 \frac{1}{z_1^2z_k^2\prod_{i\in S\setminus k}z_i^2}
 \left[-\frac{b}{a}+\frac{1}{z_1^2}+\frac{3}{z_k^2}\right].
 \label{eq:W04-k-result}
\end{equation}
Summing over $k=2,3,4$ gives
\begin{equation}
 W_{0,4}(z_1,z_2,z_3,z_4)
 =\frac{1}{z_1^2z_2^2z_3^2z_4^2}
 \left[
 -\frac{3b}{4a^3}
 +\frac{3}{4a^2}\sum_{i=1}^{4}\frac{1}{z_i^2}
 \right].
 \label{eq:W04-ab}
\end{equation}
Finally, using equation~\eqref{eq:Y-abc}, gives general equation~(\ref{eq:W04-z}).

\subsection{The case of $W_{1,2}$}
\label{app:W-12-and-TR}

\noindent For $(g,n)=(1,2)$, with $S=\{2\}$,  topological recursion procedure gives:
\begin{align}
 \omega_{1,2}(z_1,z_2)
 ={}&\operatorname*{Res}_{z=0}K(z_1,z)
 \Bigl[
 \omega_{0,3}(z,\sigma(z),z_2)
 %\nonumber\\
 %&\hspace{2.0cm}
 +B(z,z_2)\omega_{1,1}(\sigma(z))
 +\omega_{1,1}(z)B(\sigma(z),z_2)
 \Bigr]\ .
 \label{eq:omega12-recursion}
\end{align}
From equation.~\eqref{eq:omega03-a} we have:
\begin{equation}
 \omega_{0,3}(z,\sigma(z),z_2)
 =-\frac{dz^2\,dz_2}{2a\,z^4z_2^2}.
 \label{eq:omega03-zsigma-for-12}
\end{equation}
Also writing equation~\eqref{eq:omega11} as:
\begin{equation}
 \omega_{1,1}(z)=\left(\frac{A_{11}}{z^2}+\frac{B_{11}}{z^4}\right)dz,
 \qquad
 A_{11}=-\frac{b}{16a^2}\ ,
 \qquad
 B_{11}=\frac{1}{16a}\ ,
 \label{eq:omega11-ab-coeffs}
\end{equation}
since $d\sigma(z)=-dz$ we can write:
\begin{equation}
 \omega_{1,1}(\sigma(z))=-\left(\frac{A_{11}}{z^2}+\frac{B_{11}}{z^4}\right)dz.
 \label{eq:omega11-sigma}
\end{equation}
Therefore the bracket in equation~\eqref{eq:omega12-recursion} becomes:
\begin{equation}
 dz^2dz_2\left[
 -\frac{1}{2a\,z^4z_2^2}
 -\left(\frac{A_{11}}{z^2}+\frac{B_{11}}{z^4}\right)
 \left(\frac{1}{(z-z_2)^2}+\frac{1}{(z+z_2)^2}\right)
 \right]\ .
 \label{eq:omega12-bracket}
\end{equation}
This will need the expansion~(\ref{eq:Y-inverse}) of the inverse of $Y(z)$ to one order higher, 
and we also need:
\begin{equation}
 \frac{1}{(z-z_2)^2}+\frac{1}{(z+z_2)^2}
 =\frac{2}{z_2^2}\left(1+3\frac{z^2}{z_2^2}+5\frac{z^4}{z_2^4}+O(z^6)\right)\ .
 \label{eq:R-expansion}
\end{equation}
Substituting this all, expanding in even powers of $z$, and extracting the coefficient of $dz/z$ gives
\begin{align}
 W_{1,2}(z_1,z_2)
 ={}&\frac{1}{z_1^2z_2^2}
 \left[
 A_{12}+B_{12}\sum_{i=1}^{2}\frac{1}{z_i^2}
 +C_{12}\sum_{i=1}^{2}\frac{1}{z_i^4}
 +D_{12}\frac{1}{z_1^2z_2^2}
 \right]\ ,
 \label{eq:W12-shape}
\end{align}
where the coefficients  are:
\begin{equation}
A_{12}=\frac{-5ac+6b^2}{32a^4}\ ,
 \qquad
 B_{12}=-\frac{3b}{16a^3}\ ,
 \qquad
 C_{12}=\frac{5}{32a^2}\ ,
 \qquad
 D_{12}=\frac{3}{32a^2}\ .
 \label{eq:W12-abcd}
\end{equation}
Finally, substituting equation~\eqref{eq:Y-abc} into equation~\eqref{eq:W12-abcd} yields the expressions given for the general equation~(\ref{eq:W12-z}).

\section{Ref.~\cite{Johnson:2026twg}'s Closed form Formula for $W_{4,n}(\{\hat z_i\})$}
\label{app:W4n-closed-formula}

 The full closed form genus four formula, for general $(\Gamma,\{t_k\}$) is: 
  \begin{align}
  \nonumber
& W_{4,n}(\{\hat z_i\},\Gamma)={(-1)^{n}(n+5)!} (4\Gamma^2-9)(4\Gamma^2-1)\left(\prod_{i=1}^{n}\frac{1}{\hat z_i^{2}}\right)\times
\\\nonumber
&\hskip1.0cm 
\Bigg[
\frac{n+6}{185794560}\Big(
(n+7)(n+8)(12\Gamma^2-83)(4\Gamma^2-29) t_1^3\\\nonumber
&\hskip2.5cm
-6(n+7)(12\Gamma^2-115)(4\Gamma^2-25) t_1t_2
+18(4\Gamma^2-49)(4\Gamma^2-25) t_3
\Big)
\\\nonumber &\hskip1.5cm+\frac{n+6}{41287680}
\Big(
(n+7)(4\Gamma^2-29)(12\Gamma^2-83) t_1^2
-2(12\Gamma^2-115)(4\Gamma^2-25) t_2
\Big) \sum_{i=1}^{n}\frac{1}{{\hat z}_i^{2}}
\\
&\nonumber \hskip1.5cm
+\frac{n+6}{16515072} t_1(12\Gamma^2-115)(4\Gamma^2-25)  \sum_{i=1}^{n}\frac{1}{{\hat z}_i^{4}}
+\frac{n+6}{13762560} t_1(12\Gamma^2-83)(4\Gamma^2-29) \sum_{1\le i<j\le n}\frac{1}{{\hat z}_i^{2} {\hat z}_j^{2}}
\\\nonumber
&\hskip1.5cm
+\frac{1}{4718592}(4\Gamma^2-49)(4\Gamma^2-25) \sum_{i=1}^{n}\frac{1}{{\hat z}_i^{6}}
+\frac{1}{11010048}(12\Gamma^2-115)(4\Gamma^2-25) \sum_{\substack{1\le i,j\le n\\ i\ne j}}\frac{1}{{\hat z}_i^{4} {\hat z}_j^{2}}
\\
&\hskip4cm
+\frac{1}{9175040}(12\Gamma^2-83)(4\Gamma^2-29) \sum_{1\le i<j<k\le n}\frac{1}{{\hat z}_i^{2} {\hat z}_j^{2} {\hat z}_k^{2}}
\Bigg].
\label{eq:W4n-general}
\end{align}   
%\vfill\eject
\end{widetext}

\bibliographystyle{apsrev4-1}
\bibliography{references}

% at the end, hardwire the bbl file in below.

\end{document}